\pdfoutput=1

\documentclass[11pt,twoside,a4paper,cmspaper,final,collab]{cms-tdr}
\def\svnVersion{191427}\def\cmsCernNo{2012-356}\def\cmsCernDate{\today}\def\cmsMessage{Published in Physics Letters B as \href{http://dx.doi.org/10.1016/j.physletb.2013.05.040}{\doi{10.1016/j.physletb.2013.05.040.}}}

\begin{document}\cmsNoteHeader{EXO-11-095}

\hyphenation{had-ron-i-za-tion}
\hyphenation{cal-or-i-me-ter}
\hyphenation{de-vices}

\RCS$Revision: 186158 $
\RCS$HeadURL: svn+ssh://svn.cern.ch/reps/tdr2/papers/EXO-11-095/trunk/EXO-11-095.tex $
\RCS$Id: EXO-11-095.tex 186158 2013-05-15 18:40:34Z alverson $
\newlength\cmsFigWidth
\ifthenelse{\boolean{cms@external}}{\setlength\cmsFigWidth{0.48\textwidth}}{\setlength\cmsFigWidth{0.6\textwidth}}
\ifthenelse{\boolean{cms@external}}{\providecommand{\cmsLeft}{top}}{\providecommand{\cmsLeft}{left}}
\ifthenelse{\boolean{cms@external}}{\providecommand{\cmsRight}{bottom}}{\providecommand{\cmsRight}{right}}
\newcommand{\re}{\ensuremath{\cmsSymbolFace{e}}}
\renewcommand{\GeVc}{\GeV}
\renewcommand{\TeVc}{\TeV}
\renewcommand{\GeVcc}{\GeV}
\renewcommand{\TeVcc}{\TeV}
\newcommand{\intlumi}{5.0\fbinv}
\newcommand{\scalefactor}{ \ensuremath{0.98 \pm 0.03}\xspace}
\newcommand{\qWlimit}{2.38\TeV}
\newcommand{\qZlimit}{2.15\TeV}
\newcommand{\qWexplimit}{2.43\TeV}
\newcommand{\qZexplimit}{2.07\TeV}
\newcommand{\GRS}{\ensuremath{G_\mathrm{RS}}\xspace}
\newcommand{\MPl}{\ensuremath{\overline{M}_\text{Pl}}\xspace}
\cmsNoteHeader{EXO-11-095} 
\title{Search for heavy resonances in the $\PW/\cPZ$-tagged dijet mass spectrum in pp collisions at 7\TeV}

\date{\today}

\abstract{A search has been made for massive resonances decaying into a quark and a
vector boson, qW or qZ, or a pair of vector bosons, WW, WZ, or ZZ,
where each vector boson decays to hadronic final states.
This search is based on a data sample corresponding to
an integrated luminosity of 5.0\fbinv of proton-proton
collisions collected in the CMS experiment at the LHC in 2011 at
a center-of-mass energy of 7\TeV. For sufficiently heavy
resonances the decay products of each vector boson are merged into a
single jet, and the event effectively has a dijet topology. The
background from QCD dijet events is reduced using recently developed
techniques that resolve jet substructure.
A 95\% CL lower limit is set on the mass of excited quark resonances decaying into
qW (qZ) at 2.38\TeV (2.15\TeV) and upper limits are set on the cross section
for resonances decaying to qW, qZ, WW, WZ, or ZZ final states.}

\hypersetup{%
pdfauthor={CMS Collaboration},%
pdftitle={Search for heavy resonances in the W/Z-tagged dijet mass spectrum in pp collisions at 7 TeV},%
pdfsubject={CMS},%
pdfkeywords={CMS, physics, dijet, jet substructure, resonances}}

\maketitle 

\section{Introduction}\label{sec:introduction}

New resonances that decay preferentially into hadronic final states are of particular interest in a variety of scenarios for physics beyond the standard model (SM) ~\cite{Anchordoqui:2008di,Cullen:2000ef,ref_diquark,ref_qstar,Baur:1989kv,ref_axi,ref_coloron,ref_gauge,ref_rsg}.
Searches for events having a pair of hadronic jets with large invariant mass have been performed by the Compact Muon Solenoid (CMS) and ATLAS experiments at the Large Hadron Collider (LHC)~\cite{exo12094,ATLASexcitedPAS}.
In principle, these searches are also sensitive to final states that include one or two massive vector bosons $\PW/\cPZ$ because the vector bosons have large hadronic branching fractions and because their masses are much smaller than those of the hypothetical parent states, i.e. they are highly ``boosted''.
This implies that the pairs of quark jets produced by the vector boson decays merge into single $\PW/\cPZ$-jets in a real detector.
Due to the large hadronic branching fractions of the vector bosons, at the highest accessible resonance masses, a search in the fully hadronic final state can be more sensitive than searches in leptonic channels.
The sensitivity of the present large mass dijet searches is limited by the presence of background from ordinary strong interaction processes that produce pairs of quark and gluon jets.

The analysis presented here exploits the enhancement of the sensitivity of a standard dijet analysis for processes that produce $\PW/\cPZ$-jets in the final state by the application of techniques that can identify $\PW/\cPZ$-jets and suppress quark and gluon jets (``$\PW/\cPZ$-tagging'').
This CMS study is performed on pp collision data at a center-of-mass energy of 7\TeV, corresponding to an integrated luminosity of \intlumi .
We consider events with two high-transverse-momentum jets in the final state.
We identify ``subjets'' inside jets using recent
developments in the area of jet substructure~\cite{catop_cms}.
Pairs of subjets are used to explicitly reconstruct W or Z bosons,
therefore substantially suppressing backgrounds from
quantum chromodynamics (QCD) interactions.
This search follows closely the procedures of the corresponding dijet search~\cite{exo12094},
performed in the same dataset, but with strongly reduced QCD background because of the $\PW/\cPZ$-tagging.

We consider three benchmark scenarios that would produce singly or doubly tagged events:
an excited quark $\cPq^*$~\cite{ref_qstar} decaying into a quark and a W or Z boson;
a Randall--Sundrum (RS) graviton \GRS~\cite{rs1} decaying to WW or ZZ;
and a heavy partner of the SM W boson $\PWpr$ which decays to WZ~\cite{ref_gauge}.
The most stringent limits on the $\cPq^*$
model have been set in dijet resonance searches at the LHC by considering
the $\cPq\cPg$ final state~\cite{exo12094} or
inclusively all-hadronic final
states~\cite{ATLASexcitedPAS}.  The most stringent
lower limit (at 95\% CL) on the $\cPq^*$ mass to date is
3.3\TeVcc~\cite{exo12094}.  Specific searches for the $\cPq\PW$ and
$\cPq\cPZ$ final states have previously been reported at the
Tevatron~\cite{CDFexcitedPAPER,D0excitedPAPER}, which exclude resonances decaying to $\cPq\PW$ or
$\cPq\cPZ$ with masses up to 540\GeVcc, and at the LHC~\cite{CMSqZPAS},
which extends the mass exclusion of $\cPq\cPZ$ resonances up to 1.94\TeVcc .  For the \GRS, there
are phenomenological models favoring the decay of the \GRS into
vector bosons rather than photons or
fermions~\cite{GravitonWWZZ1,GravitonWWZZ2,GravitonWWZZ3}.
In particular, the ZZ final state has been explored
experimentally~\cite{CMSZZPAS2,ATLASZZPAPER,CDFZZPAPER},
setting lower limits on the \GRS mass as a function of the
coupling parameter $k/\MPl$,
where $k$ is the curvature of the warped space and $\MPl$ the reduced Planck mass ($\MPl \equiv M_\text{Pl}/\sqrt{8\pi}$).
For the $\PWpr$, the most stringent
limits are reported in searches with leptonic final
states~\cite{CMSwprimePAS,ATLASwprimePAPER},
and the current lower limit on the $\PWpr$ mass is 2.5\TeVcc.
The limit varies by 0.1\TeVcc, depending on the chirality of the $\PWpr$ couplings.
Specific searches in the WZ final state have also been
reported~\cite{CMSwprimeWZPAS,ATLASwprimeWZPAS} setting a lower limit of 1.1\TeVcc .

This paper is organized as follows.
First, the CMS detector, and the simulated and collision data samples on which the analysis is based are briefly described.
Then the event reconstruction, and selection are detailed, and the $\PW/\cPZ$-tagging technique is described.
The following section describes the modeling of detector acceptances and signal efficiencies as well as the validation of the $\PW/\cPZ$-tagging techniques with data.
After this follows the description of the modeling of the background, the systematic uncertainties and the limit setting procedure.
Finally the results and conclusions are presented.
\section{CMS detector}
\label{sec:cms_detector}

The CMS detector~\cite{:2008zzk} is well-suited to the reconstruction of
hadronic jets because it incorporates finely segmented electromagnetic
and hadronic calorimeters, and a charged-particle tracking system.
Charged particles are reconstructed in the inner tracker, which is
immersed in a 3.8\unit{T} axial magnetic field. The inner tracker consists
of three cylindrical layers and two endcap disks at each end of silicon pixel detectors, and ten
barrel layers and twelve endcap disks at each end of silicon strip detectors.  This
arrangement results
in full azimuthal coverage ($0 \le \phi \le 2 \pi$) within $|\eta| < 2.5$, where $\eta$
is the pseudorapidity defined as $\eta = -\ln[\tan(\theta/2)]$.
CMS uses a polar coordinate system, with the
$z$ axis coinciding with the beam axis;
$\theta$ is the polar angle defined with respect to the positive $z$ axis.
Muons are measured in gas-ionizing detectors embedded in the steel return yoke.
A lead-tungstate crystal electromagnetic calorimeter (ECAL) up to $|\eta|=3$ and
a brass/scintillator hadronic calorimeter (HCAL) up to $|\eta|=5$ surround the tracking
volume and allow photon, electron, and jet reconstruction.
The ECAL and HCAL cells are grouped into towers projecting radially
outward from the interaction region.  In the central region ($|\eta|<1.74$)
the towers have dimensions $\Delta\eta = \Delta\phi = 0.087$;
at higher $|\eta|$, the $\Delta\eta$ and $\Delta\phi$ widths increase.
For optimum jet reconstruction, the tracking and calorimeter information is combined
in an algorithm called particle flow~\cite{particleflow}, which is described
below.
\section{Simulated and collision data samples} \label{sec:data_and_mc_samples}

The sample of proton-proton collision data at $\sqrt{s}=7$\TeVcc, corresponding to an integrated luminosity of
\intlumi, was collected in 2011.
The events were collected using the logical ``or'' of a set of
triggers based on requirements on $H_{\text{T}} = \sum_{\text{jets}} \pt$
($\pt$ is the transverse momentum of a jet) and
the invariant mass of the two highest $\pt$ jets in an event,
whose thresholds were raised progressively to cope with an increase in the
peak luminosity during 2011.

Data are compared to Monte Carlo (MC) simulations of the QCD background
generated using both \PYTHIA6.424~\cite{pythia} and \HERWIG{++} 2.4.2~\cite{herwig}.
\PYTHIA6 is used with CTEQ61L~\cite{cteq} and \HERWIG{++} with MRST2001~\cite{mrst} parton distribution functions.
Tune Z2 (identical to tune Z1~\cite{bib_tunez1} except that Z2 uses the CTEQ6L PDF while Z1 uses CTEQ5L) is used with \PYTHIA6,
while the tune version 23~\cite{herwig} is used with \HERWIG{++}.
In this analysis, the background shape is modeled from the data themselves.
Therefore, the analysis depends on QCD simulation only to provide guidance and cross checks.

The sensitivity of the event selection to the benchmark processes is evaluated
using simulated samples of events from excited quarks, RS gravitons, and $\PWpr$ production and decay models.
The process $qg \to \cPq^* \to \PW/\cPZ + \text{jet}$ is generated using \PYTHIA6 assuming the coupings to the SU(2), U(1) and SU(3) groups are $f=f'=f_s=1$ for the production and decay of the $\cPq^*$.
The process $\GRS \to \PW\PW/\cPZ\cPZ$ is generated using \HERWIG{++}
and its cross section is taken from \PYTHIA6.
While \HERWIG{++} contains a more detailed description of the angular distributions than \PYTHIA6 for this process~\cite{resonanceshape},
the cross section is taken from \PYTHIA6 which has been used as a reference model in related analyses~\cite{CMSZZPAS2}.
RS graviton production is studied with $k/\MPl=0.1$, which determines a resonance
width of about 1\% of the resonance mass which is about a factor 5 smaller than the experimental resolution for dijets.
This width is much smaller than suggested by the model in Ref.~\cite{GravitonWWZZ1}, which predicts resonance widths of the order of the experimental resolution, allowing for interpretation in this model only approximately.
The process $\PWpr \to \PW\cPZ$ is generated using \PYTHIA6 with Standard Model $V-A$ couplings and without applying k-factors.
All Monte Carlo events are passed through the CMS detector simulation based on \GEANTfour~\cite{refGEANT}.
\section{Event reconstruction and selection}\label{sec:analysis}

Events are reconstructed using the particle flow
algorithm, which attempts to identify and measure all the stable
particles in a collision by combining information from all the
subdetectors. This algorithm categorizes all particles into five types:
muons, electrons, photons, charged and neutral hadrons. The resulting
particle flow candidates are passed to the anti-\kt~\cite{ktalg} and Cambridge-Aachen
(CA)~\cite{CAaachen,CAcambridge} jet clustering algorithms, as
implemented in \textsc{FastJet}~\cite{fastjet1,fastjet} to create jets.
A distance parameter of size $R=\sqrt{(\Delta
  \eta)^2 + (\Delta\phi)^2}=0.8$ is used for the CA algorithm, while
$R=0.5$ is used for the anti-\kt algorithm. While the
anti-\kt jets are used to select events and reconstruct
the dijet invariant mass $m_{jj}$, the CA jets are used to identify
events containing hadronically decaying W or Z bosons.
This choice has been made because the CA algorithm was found to be more efficient (for the same mistag rate) at
finding hard subjets within the jets in simulation-based studies~\cite{catop_cms},
while the anti-\kt jets have the best energy calibration.

Events must have at least one reconstructed vertex within
$\abs{z} < 15\unit{cm}$, to suppress backgrounds solely triggered by calorimeter noise.
The primary vertex is defined as the vertex with highest sum of squared track transverse momenta ($\pt^2$).
Charged particles not originating from it are removed from the inputs to the jet clustering algorithms.
This requirement removes particles which arise from additional pp interactions
in the same pp bunch crossing (pileup interactions).
An event-by-event jet-area-based correction~\cite{jetarea_fastjet,jetarea_fastjet_pu,JME-JINST} is applied to remove the remaining
pileup energy which is due to neutral particles originating from
the other vertices.
The pileup-subtracted jet four momenta are finally corrected
to account for the difference between the measured and
true responses to hadrons~\cite{JME-JINST}.
When jets are decomposed into subjets, as described later,
the energy estimate relies on the calibrated
reconstructed input particles without further corrections.

Events are initially selected by requiring that they have at least
two anti-\kt jets with $\pt > 30\GeVc$ and $|\eta| < 2.5$.
The two highest-$\pt$ jets are required to
have a pseudorapidity separation $|\Delta\eta|<1.3$ to reduce the QCD dijet background~\cite{CMSexcitedPAPER}.
Finally, the dijet invariant mass is required to be larger than 890\GeVcc.
This threshold is defined by the triggers, which were found to be 99\% efficient for dijet events with masses above this threshold.

In events passing this selection, ``boosted'' (high $\pt$)
hadronically decaying W or Z bosons are identified with a $\PW/\cPZ$-tagging
algorithm using jet pruning~\cite{topwtag_pas}, a technique which removes the
softest components of the jets.
In the jet pruning technique~\cite{jetpruning1,jetpruning2}, a jet is reclustered using all the particles used to build a CA jet,
ignoring in each recombination step the softer ``protojet'' if the recombination is softer than a given threshold $z_{\text{cut}}=0.1$ or forms an angle $\Delta R$
wider than $D_{\text{cut}}=0.5 m^\text{orig}/\pt^\text{orig}$ with respect to the previous recombination step,
where $m^\text{orig}$ and $\pt^\text{orig}$ are the mass and transverse momentum of the original CA jet.
The hardness of a recombination $z$ is defined as $z=\min(\pt^i,\pt^j)/\pt^p$, where $\pt^i$ and $\pt^j$ are the $\pt$ of the two protojets to be combined and $\pt^p$ is the $\pt$ of the combined jet.
The following selection is
then applied to the pruned jets to identify jets from hadronic $\PW/\cPZ$
decays by exploiting the variables used in Ref.~\cite{boostedhiggs}.
The total pruned jet mass $m_\text{jet}$ must satisfy $70\GeVcc < m_\text{jet} < 100\GeVcc $.
Two subjets are obtained by undoing the last clustering iteration of the pruned jet clustering.
The ratio of masses of the highest mass subjet ($m_1$) and the total pruned jet mass is defined as the \textit{mass drop} $\frac{m_1}{m_\text{jet}}$.
To discriminate against QCD jets, the mass drop is required to satisfy $\frac{m_1}{m_\text{jet}} < 0.25$.
These criteria are designed to select W and Z candidates in which the subjets are similar in energy and mass.

\begin{figure}[thb]
\centering
     \includegraphics[width=\cmsFigWidth]{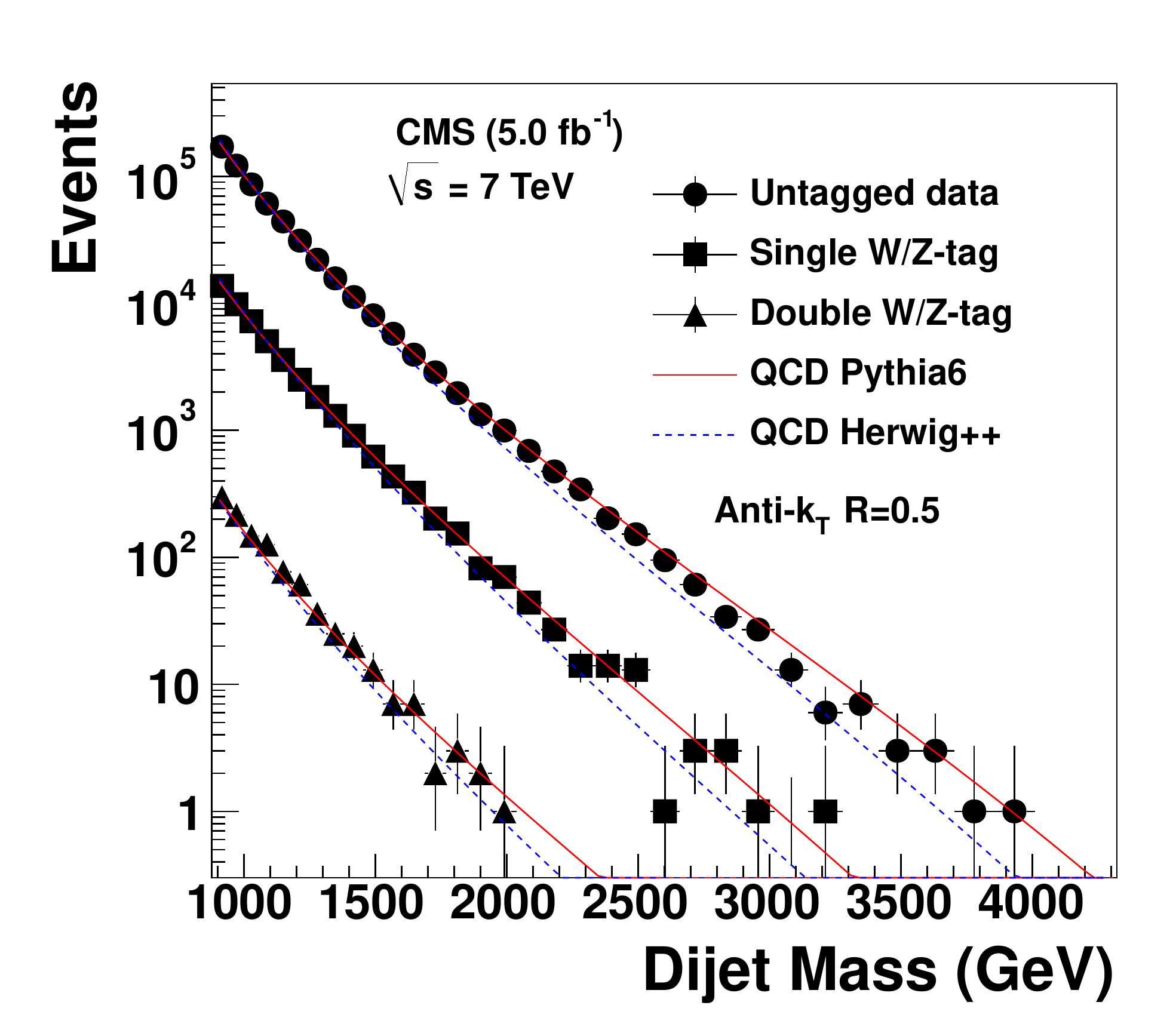}
\caption[Invariant Mass]{Comparisons of the dijet invariant mass
  distributions between data and Monte Carlo (\PYTHIA6 and
  \HERWIG{++}) simulations.  The three sets of lines correspond to the
  inclusive dijet category (no $\PW/\cPZ$-tag required), single $\PW/\cPZ$-tagged,
  and double $\PW/\cPZ$-tagged events.  The simulations are normalized to the number of
  data events in each category.}
  \label{fig:mjj}
\end{figure}

Comparisons of the dijet invariant mass distributions for untagged, single-tagged,
and double-tagged event samples are shown in Fig.~\ref{fig:mjj}.
The data are shown as solid points and the \PYTHIA6 and \HERWIG{++} simulations are shown as solid
red and dashed blue curves, respectively.
The simulations are normalized to the number of data events
in each category and the shapes are compared;
the agreement of the normalization driven by the $\PW/\cPZ$-tagging efficiency is discussed in the next section.
The \PYTHIA6 prediction is found to
agree with the data while the \HERWIG{++} prediction decreases more steeply with mass.
However, no systematic uncertainties are taken into
account and only the dominant background from QCD interactions is considered.

\section{Signal characterization}
\label{sec:signal}

A search for dijet resonances corresponding to several benchmark physics models is performed.
Using the $\PW/\cPZ$-tagging algorithm, both single $\PW/\cPZ$-tag and double $\PW/\cPZ$-tag events are examined.
The signals that would be produced by the benchmark physics models have different characteristics that are described below.

\begin{figure}[th!b]
\begin{center}
\includegraphics[width=0.46\textwidth]{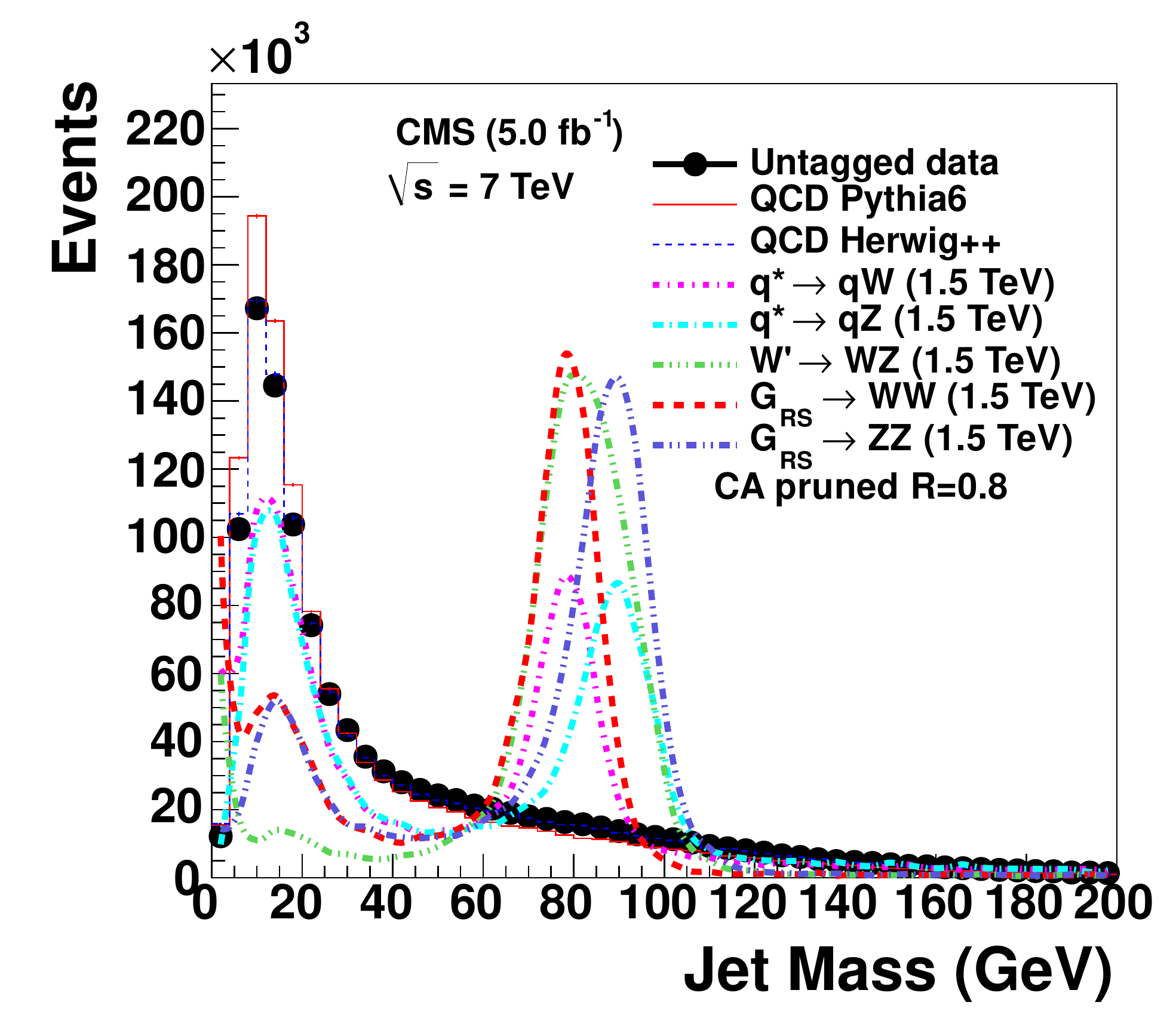}
\includegraphics[width=0.46\textwidth]{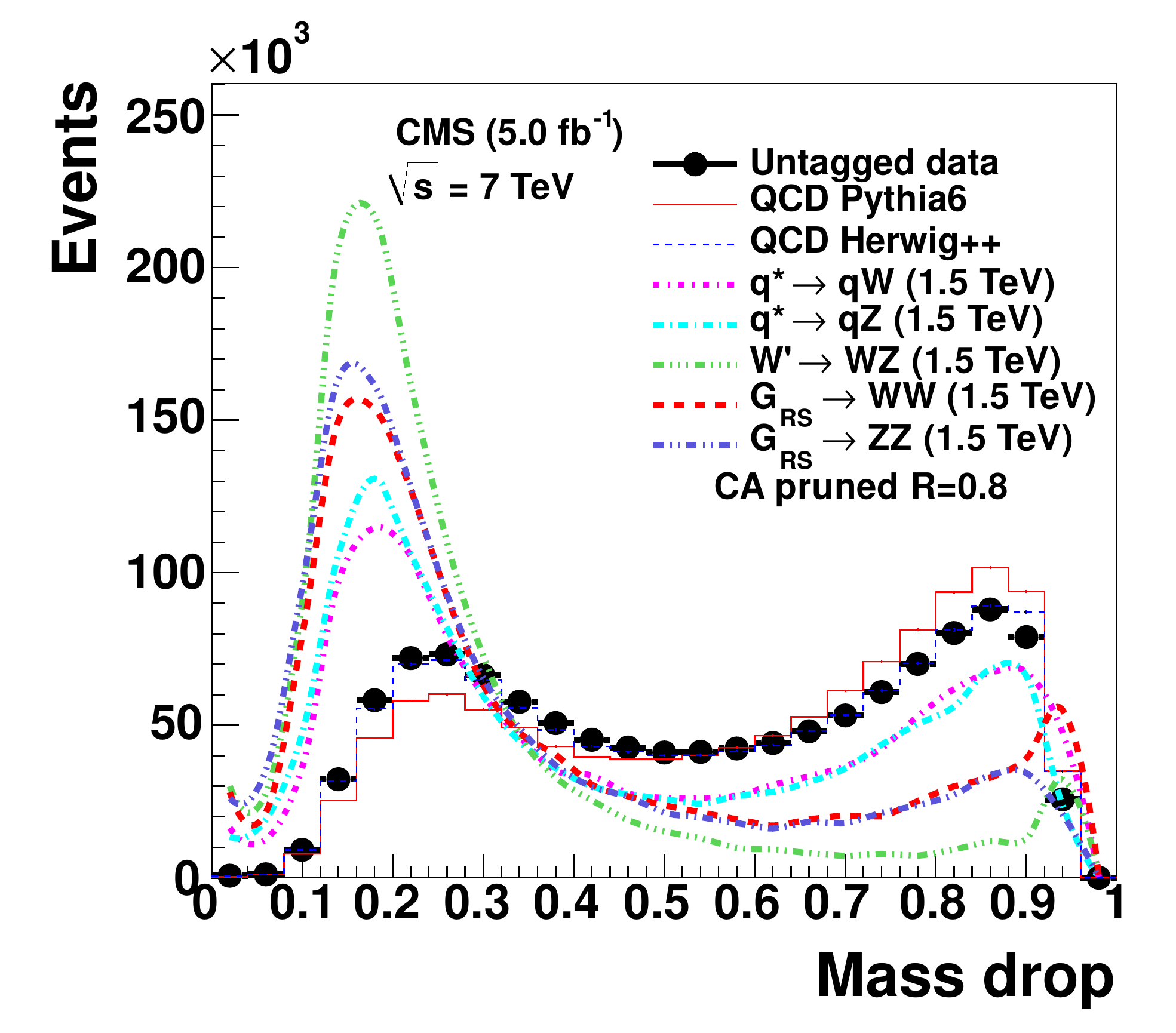}
\end{center}
\caption{Pruned jet mass (\cmsLeft) and mass drop (\cmsRight) in data,
  signal and background simulations.
  The signal simulation distributions are plotted as smooth curves connecting the
  histogram entries (using the same binning as the data distribution).
  All simulation distributions have been scaled to match the number of data events.
}
\label{fig:taggingvariables}
\end{figure}

The pruned jet mass and mass drop distributions in data,
signal, and background simulations are shown in Fig.~\ref{fig:taggingvariables}.  The
discriminating power of the pruned jet mass and mass drop for the
different signals is evident.
In both the pruned jet mass and the mass drop distributions, small differences may be seen between the results obtained with \HERWIG{++} ($\PW\PW$, $\cPZ\cPZ$) and  \PYTHIA6 ($\PW\cPZ$, $\cPq\PW$, $\cPq\cPZ$), which arise from differences in the showering and hadronization models used by these generators.
This effect is taken into account in the estimate of the systematic uncertainties on the tagging efficiency, as described below.

\begin{figure}[thb]
\begin{center}
\includegraphics[width=0.48\textwidth]{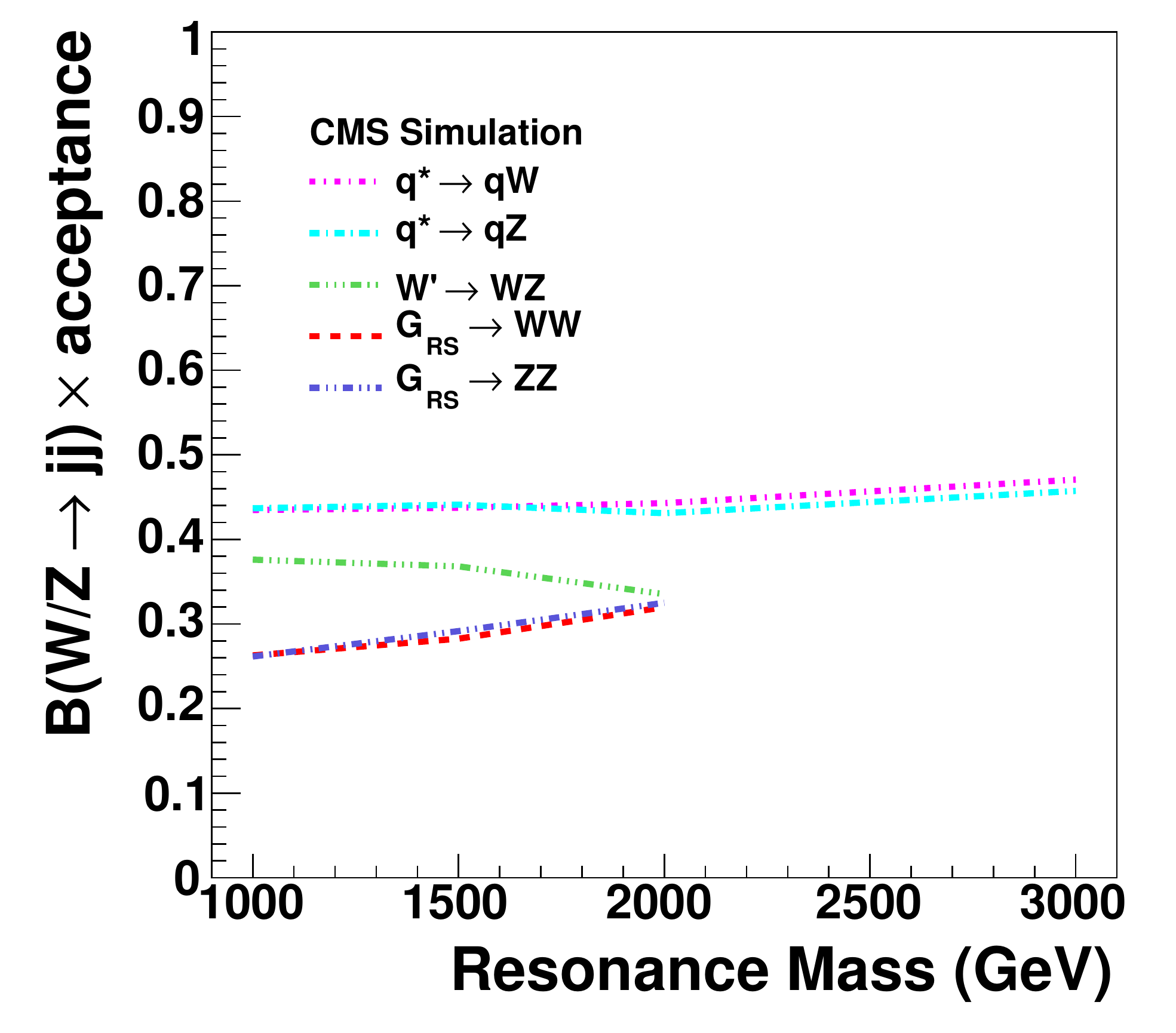}
\end{center}
\caption{The branching fraction into dijet final states B($\PW/\cPZ \to \text{jets}$)
  times angular acceptance ($|\eta| < 2.5$, $|\Delta\eta|<1.3$).
  The $\PW/\cPZ$-tagging efficiencies are excluded from the acceptance.
}
\label{fig:acceptance}
\end{figure}

\begin{figure}[htb]
\begin{center}
\includegraphics[width=0.46\textwidth]{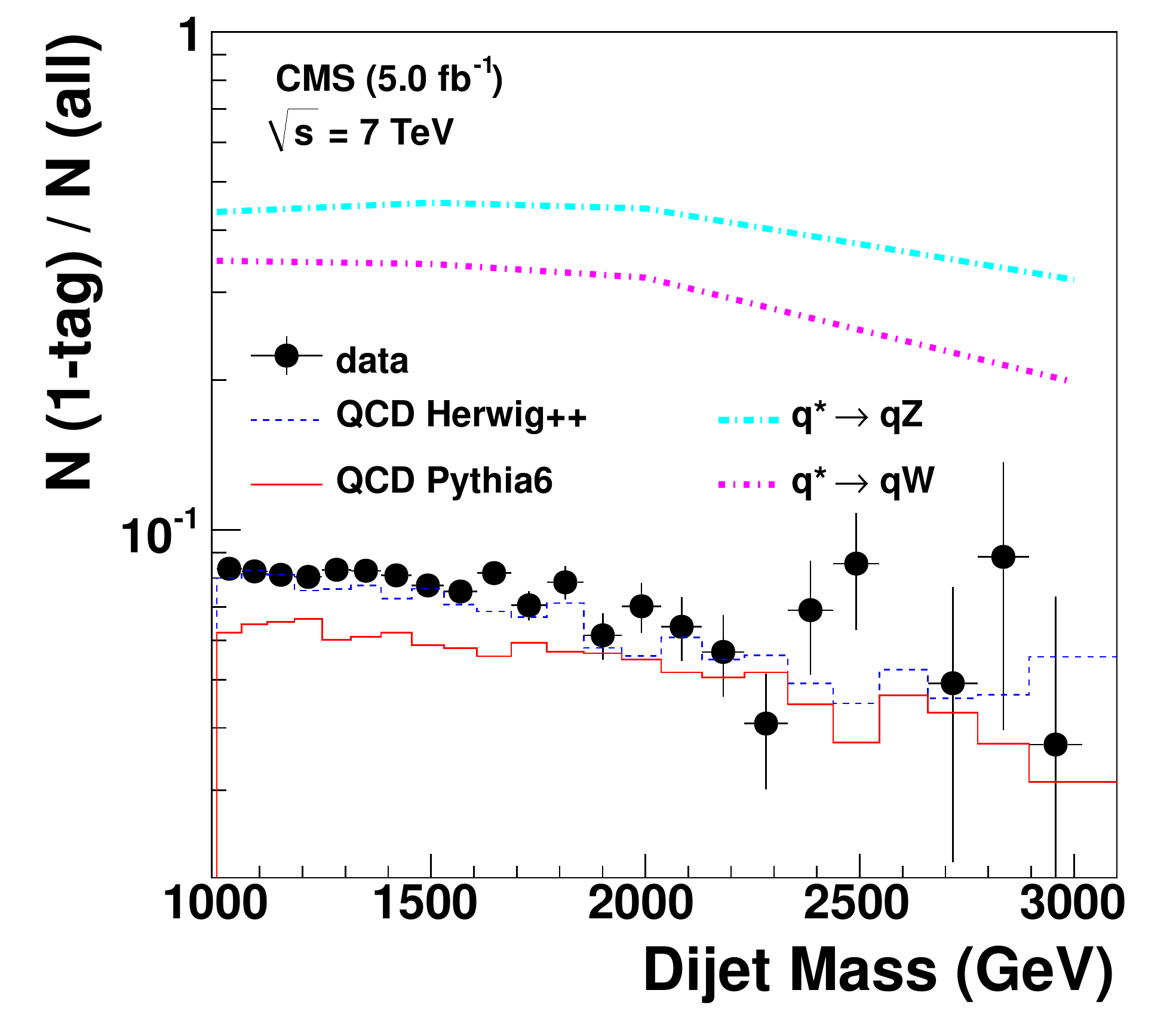}
\includegraphics[width=0.46\textwidth]{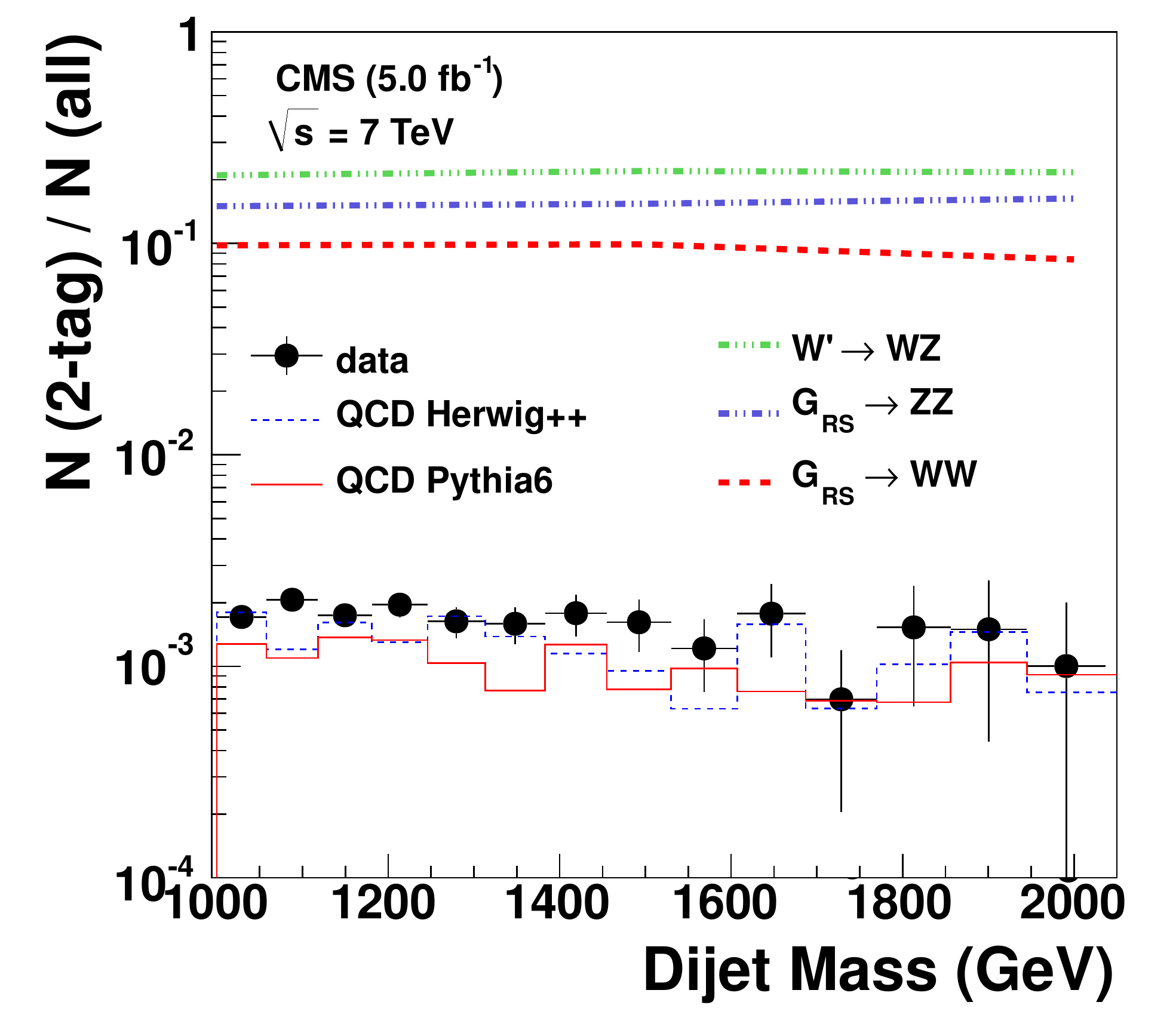}
\end{center}
\caption{Efficiency of requiring 1 $\PW/\cPZ$-tag (\cmsLeft) and 2 $\PW/\cPZ$-tags (\cmsRight) in
  signal and background simulations, and in data for events passing the angular acceptance
  requirement ($|\eta| < 2.5$, $|\Delta\eta|<1.3$).
}
\label{fig:efficiencies}
\end{figure}

\begin{figure}[htb]
\begin{center}
\includegraphics[width=0.48\textwidth]{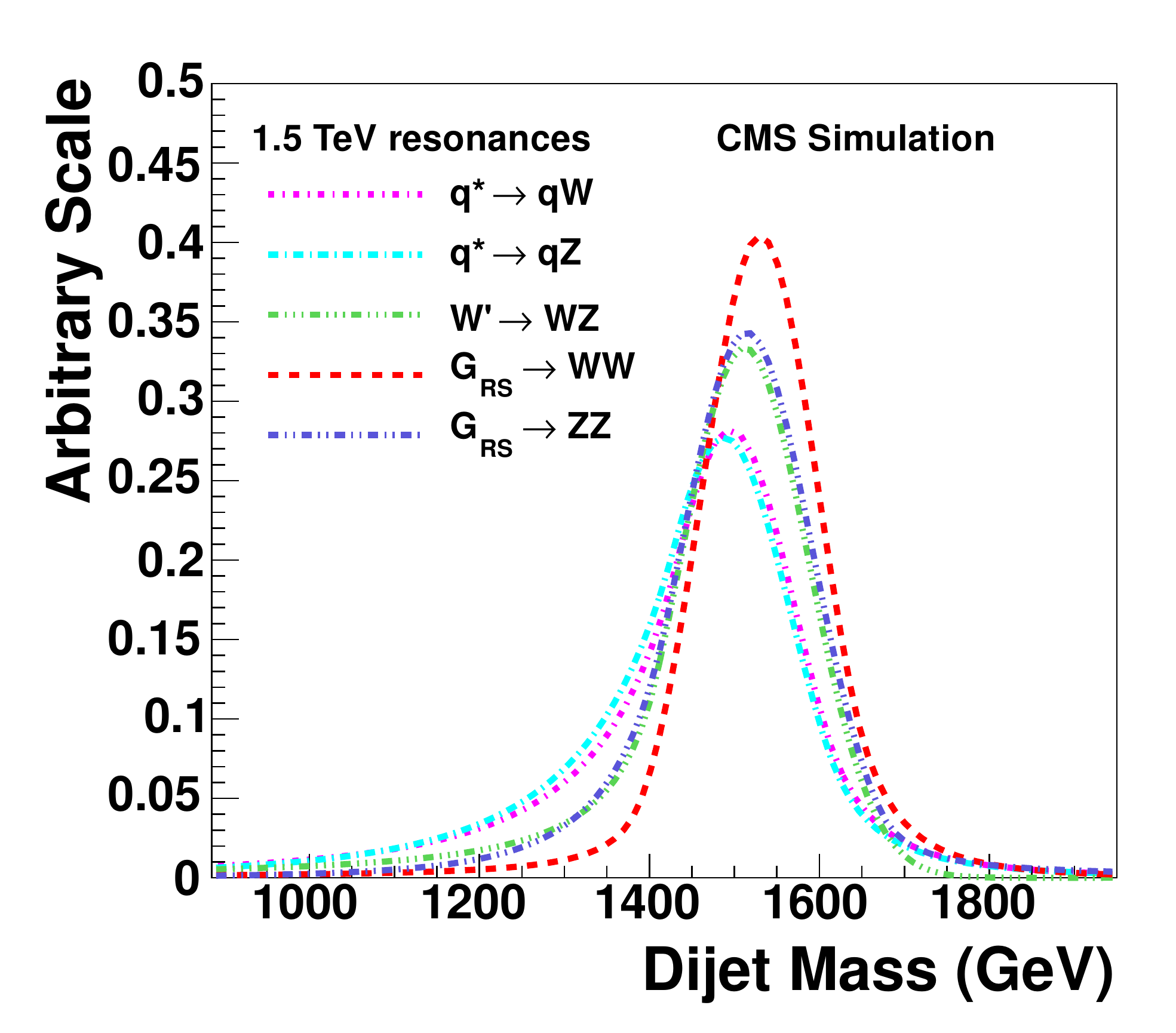}
\end{center}
\caption{The signal dijet invariant mass distributions for 1.5\TeVcc
  $\GRS\to \PW\PW$, $\GRS\to \cPZ\cPZ$, $\PWpr \to \PW\cPZ$,
  $\cPq^*\to \cPq\PW$, and $\cPq^*\to \cPq\cPZ$ resonances,
  computed using anti-\kt jets with $R=0.5$.
  Crystal-Ball functions fit to the simulated distributions are shown.
  The distributions in the plot are scaled to the same integral.
}
\label{fig:resonanceshape}
\end{figure}

The acceptance, defined as the product of signal branching fraction into dijet final states B($\PW/\cPZ \to \text{jets}$)
times angular acceptance ($|\eta| < 2.5$, $|\Delta\eta|<1.3$), is shown in
Fig.~\ref{fig:acceptance}. Each model relevant for the singly (doubly) tagged data analysis
is shown in the dijet invariant mass range up to 2\TeV(3\TeV).
The fraction of events that
produce dijet events, which have survived the kinematic selection,
is between $26\%$ and $47\%$.  This fraction includes
also the branching fraction of $\PW/\cPZ$ decaying into objects which are
reconstructed as jets.
The different behaviour of the acceptance for $\PWpr$ and \GRS at low dijet masses is due to different
angular distributions generated by \PYTHIA6 and \HERWIG{++}.

The $\PW/\cPZ$-tagging efficiency, which is not part of the acceptance, is shown for signal and background events
in Fig.~\ref{fig:efficiencies}.  The signal efficiency, determined
from the simulation, is found to be between 20\% and
45\% (8\% and 22\%) for single (double) $\PW/\cPZ$-tagged signals.
The $\PW$-tagging efficiency is larger than the $\cPZ$-tagging efficiency due to the choice of the jet mass window cut,
which rejects a larger fraction of W bosons by requiring $m_\text{jet}>70$\GeV.
The simulation modeling of the signal efficiency is cross-checked by measuring the
$\PW/\cPZ$-tagging efficiency in semileptonic $\ttbar$ data and by comparing it
with the same efficiency obtained using the same procedure for
$\ttbar$ simulation generated with {\MADGRAPH 4.4.12}~\cite{madgraph}
and showered with \PYTHIA6.
We follow the same procedure as described in Ref.~\cite{CMSttbar}.
The ratio of the two efficiencies
results in a scale factor of \scalefactor
which is then applied to the efficiencies for
signals in the dijet data.  The uncertainties on the scale factor are
propagated into the systematic uncertainties on the overall signal
efficiency.

As described above, the production and decay of the \GRS is modeled with \HERWIG{++}.
A difference of up to 18\% on the double-tag
efficiency in the RS graviton $\PW\PW/\cPZ\cPZ$ signal simulations between \PYTHIA6 and \HERWIG{++} is observed.
The difference can be attributed in equal parts to the different showering algorithms and the hadronization algorithms.
The different underlying event modeling has only a small impact of $<$1\%.
This discrepancy is accounted for as a systematic uncertainty on the double-tag efficiency.
For the single-tag efficiency a 9\% uncertainty is assigned based on
the double-tag efficiency uncertainty of 18\% assuming
an identical difference in tagging efficiency for the two vector bosons
in the case of 100\% efficiency, representing an upper limit on this uncertainty.

The effect of pileup on the $\PW/\cPZ$-tagging efficiency was also checked.
Because of the rejection of charged particles not originating from the primary vertex
and the application of pruning, the pileup dependence is weak and the uncertainty of the
modeling of the pileup distribution is less than 2\%.

The dijet mass dependence of the $\PW/\cPZ$-tagging efficiency for
background events shown in Fig.~\ref{fig:efficiencies}
is adequately described by the simulation.
Therefore, no additional systematic uncertainty is assigned on the
dijet mass dependence of the modeling of the $\PW/\cPZ$-tagging in simulation.

Figure~\ref{fig:resonanceshape} shows the signal shapes for $\GRS \to
\cPZ\cPZ/\PW\PW$, $\PWpr \to \PW\cPZ$, and $\cPq^* \to \cPq\PW/\cPq\cPZ$,
all of which correspond to a resonance mass of 1.5\TeV.
The differences for the different models are to a large extent due to the different tagging efficiencies for W and Z and to a smaller extent to differences in the models in \PYTHIA6 and \HERWIG{++}.
The lower cut of 70\GeV on the jet mass in the $\PW/\cPZ$-tag biases the resonance peak for $\PW\PW$, $\PW\cPZ$ and $\cPq\PW$ towards higher masses, especially when two tags are applied on the $\PW\PW$ sample. We have checked that this behavior is reproduced in both \PYTHIA6 and \HERWIG{++}.
The difference in resolution between the singly tagged and doubly tagged resonance shapes is also due to this bias of the extra $\PW/\cPZ$-tag requirement.
Resonance shapes were simulated at masses of 1, 1.5, 2 and 3\TeV and a linear interpolation was used to obtain the shapes at intermediate masses.

\begin{figure}[th!b]
\begin{center}
\includegraphics[width=0.48\textwidth]{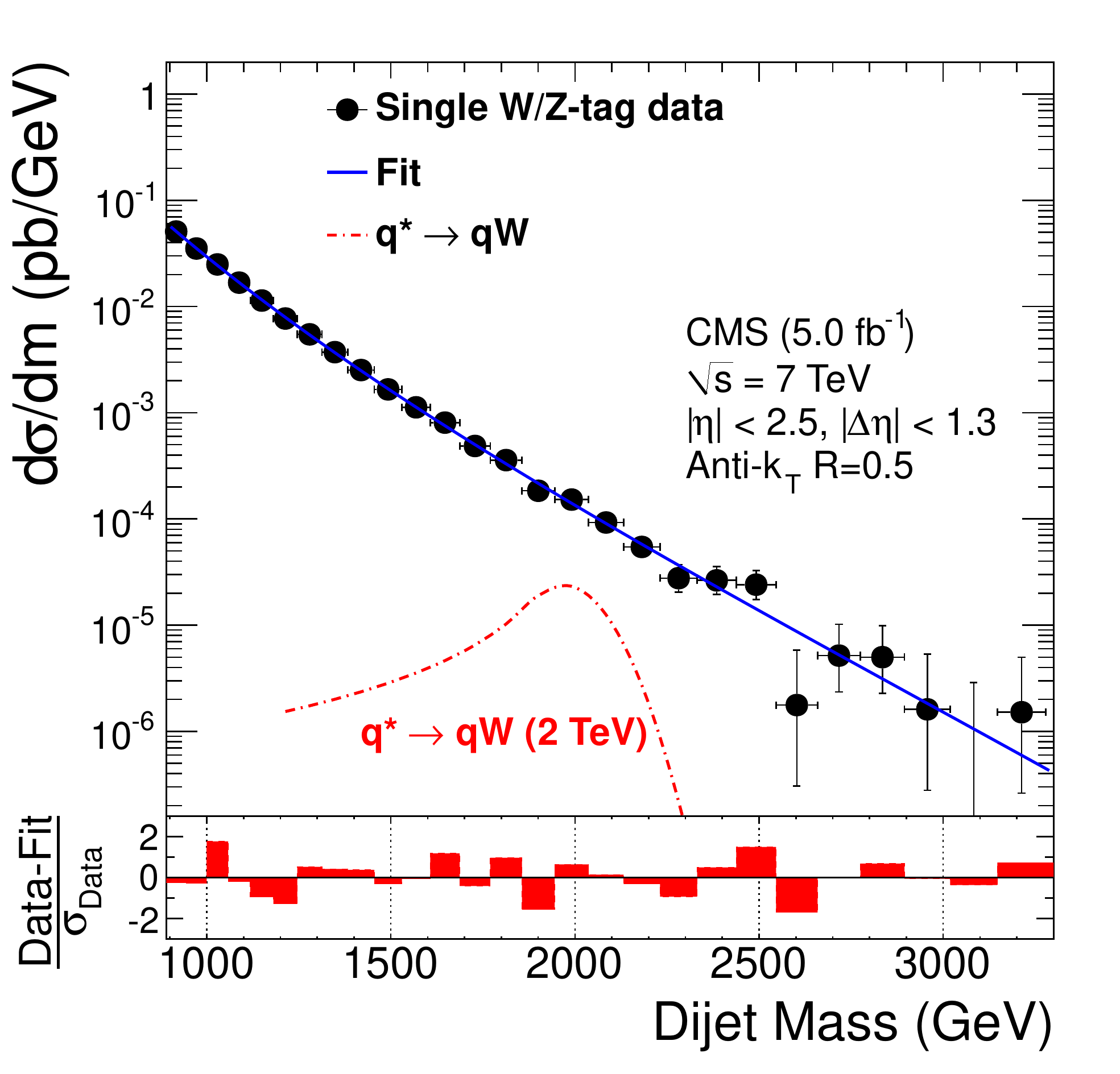}
\includegraphics[width=0.48\textwidth]{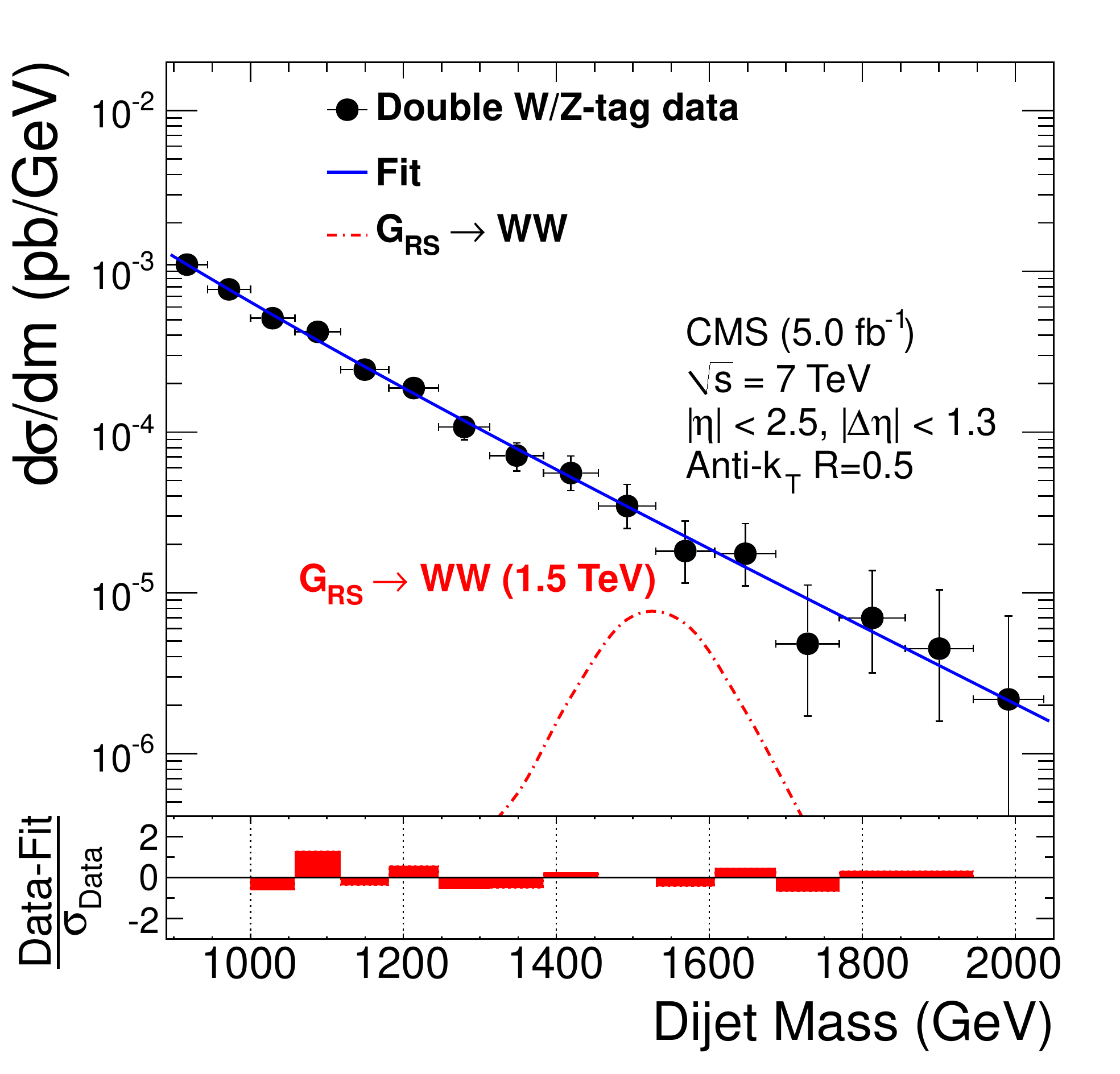}
\end{center}
\caption{The single (\cmsLeft) and double (\cmsRight) $\PW/\cPZ$-tagged $m_{jj}$
  distributions (points) in data fitted with the QCD background parametrization (solid
  curve).  For the double $\PW/\cPZ$-tagged distribution $P_3=0$ is assumed in
  Eq.~(\ref{eqParam}).  Signal shape distributions for $\cPq^* \to \cPq\PW$ and
  $\GRS \to \PW\PW$ with arbitrary cross sections are also shown. The bottom panes show
  the corresponding pull distributions ($\frac{\text{Data}-\text{Fit}}{\sigma_{\text{Data}}}$).
  There are no data events with dijet masses larger than the range of these plots.}
\label{fig:singleVtagBG}
\end{figure}

\section{Background shape parametrization}
\label{sec:background}

The shape of the QCD background in the dijet spectrum is modeled using a
simple parametrization which has been successfully deployed in previous
searches~\cite{cmsdijet}.  The background
model is given by:
\begin{equation}
\frac{\rd\sigma}{{\rd}m} =
\frac{P_{0} (1 - m/\sqrt{s})^{P_{1}}}{(m/\sqrt{s})^{P_{2} + P_{3} \ln
(m/\sqrt{s})}} \ ,
\label{eqParam}
\end{equation}
\noindent where $m$ denotes the dijet mass and $\sqrt{s}$ the pp center of mass energy.
$P_0$ acts as a normalization parameter for the probability
density function, and $P_1$, $P_2$, $P_3$ describe its shape.
For the single $\PW/\cPZ$-tagged analysis, all parameters are free to float
in the fit.  For the double $\PW/\cPZ$-tagged analysis,
$P_3$ is not needed as suggested by a Fisher F-test~\cite{Ftest} and
a simpler parametrization with $P_3$ fixed to 0 is used.

Figure \ref{fig:singleVtagBG} shows the dijet mass spectra from
single and double $\PW/\cPZ$-tagged data fitted to Eq.~(\ref{eqParam}) and the corresponding pull
distributions, demonstrating the agreement between the background-only
probability density function and the data.

Since no sizeable deviation from the background-only hypothesis is seen,
exclusion limits are set on the product of cross section, acceptance, and branching fraction for
the five considered final states: qW, qZ, WW, WZ, and ZZ.

\section{Systematic uncertainties}
\label{sec:systematics}
The sources of systematic uncertainties are summarized as follows.
The only background-related systematic uncertainty is the choice of background parametrization which is discussed in Section~\ref{sec:statistics}.
The leading signal-related systematic uncertainties are the $\PW/\cPZ$-tagging efficiency (Section~\ref{sec:signal}), jet energy scale (JES), jet energy resolution (JER), and luminosity measurement.
Because the trigger and reconstruction efficiencies are larger than 99\% in the relevant dijet mass range,
the uncertainties associated with these efficiencies are negligible.

In the jet \PT- and $\eta$-regions considered in this analysis, the JES
has an uncertainty of 2-3\%~\cite{JME-JINST}.  The
\PT- and $\eta$-dependent uncertainty is propagated to an uncertainty on the
reconstructed dijet invariant mass of 2.2\%, which is approximately mass independent.
The effect of the JES uncertainty on the calculation of the limits
is estimated by varying the resonance dijet mass in the statistical analysis.
The JER is known to a precision of 10\% and its tails are in
agreement between data and simulation~\cite{JME-JINST}.
The effect of the JER uncertainty on the calculation of the limits is estimated
by varying the reconstructed resonance width in the statistical analysis.
The luminosity has an uncertainty of 2.2\%~\cite{SMP-12-008-PAS},
which is also taken into account in the statistical analysis.

\section{Limit setting procedure}
\label{sec:statistics}

For setting upper limits on the resonance production cross section
a Bayesian formalism with uniform prior for the cross section is used,
following the procedure used in Ref.~\cite{cmsdijet}.
The binned likelihood, $L$, can be written as:
\begin{equation}
L = \prod_{i} \frac{\mu_{i}^{n_{i}}\re^{-\mu_{i}}}{n_{i}!} \ ,
\end{equation}
where
\begin{equation}
{\mu_{i}} = {\alpha}{N_{i}(S)} + {N_{i}(B)} \ ,
\label{function}
\end{equation}
$n_i$ is the observed number of events in the $i^{th}$ dijet mass bin,
$N_i(S)$ is the expected number of events from the signal in the $i^{th}$ dijet
mass bin, $\alpha$ scales the signal amplitude, and
$N_i(B)$ is the expected number of events from background in the
$i^{th}$ dijet mass bin.
The background $N_i(B)$ is estimated as the background component
of the best 5(4)-parameter fit of equation~\ref{function} to the singly (doubly) tagged data points.
The signal is not restricted to be positive for the background estimate fit
although it is restricted in the Bayesian prior for the signal.
A flat prior in $\alpha$, which
is the same as a flat prior in the resonance production cross section, is assumed.

The dominant sources of systematic uncertainty (the jet energy
scale, the jet energy resolution, the integrated luminosity, and the
$\PW/\cPZ$-tagging efficiency) are considered as nuisance parameters associated to log-normal priors.
The uncertainty on the background shape is taken into account with nuisance parameters associated to Gaussian priors
representing variations of the fit parameters along the eigenvectors of their correlation matrix.
The systematic uncertainties are accounted for using a fully Bayesian
treatment and integrating the likelihood over nuisance parameters.

The 95\% confidence level (CL) upper limit $\sigma_S$ is calculated from
the normalized posterior probability density $P_\text{post}$ as follows:
\begin{equation}
\int_{0}^{\sigma_S} {P_\text{post}(\sigma) }\,\rd\sigma=0.95 \ .
\end{equation}

This method of using the data first to constrain the background fit and second to extract the limit
induces a bias in the coverage of the limits.
The actual coverage is reduced to 93.6\% (94.3\%) at a WW (qW) signal mass of 1200\GeV (1800\GeV).

\begin{figure*}[h!tpb]
\begin{center}
\includegraphics[width=0.43\textwidth]{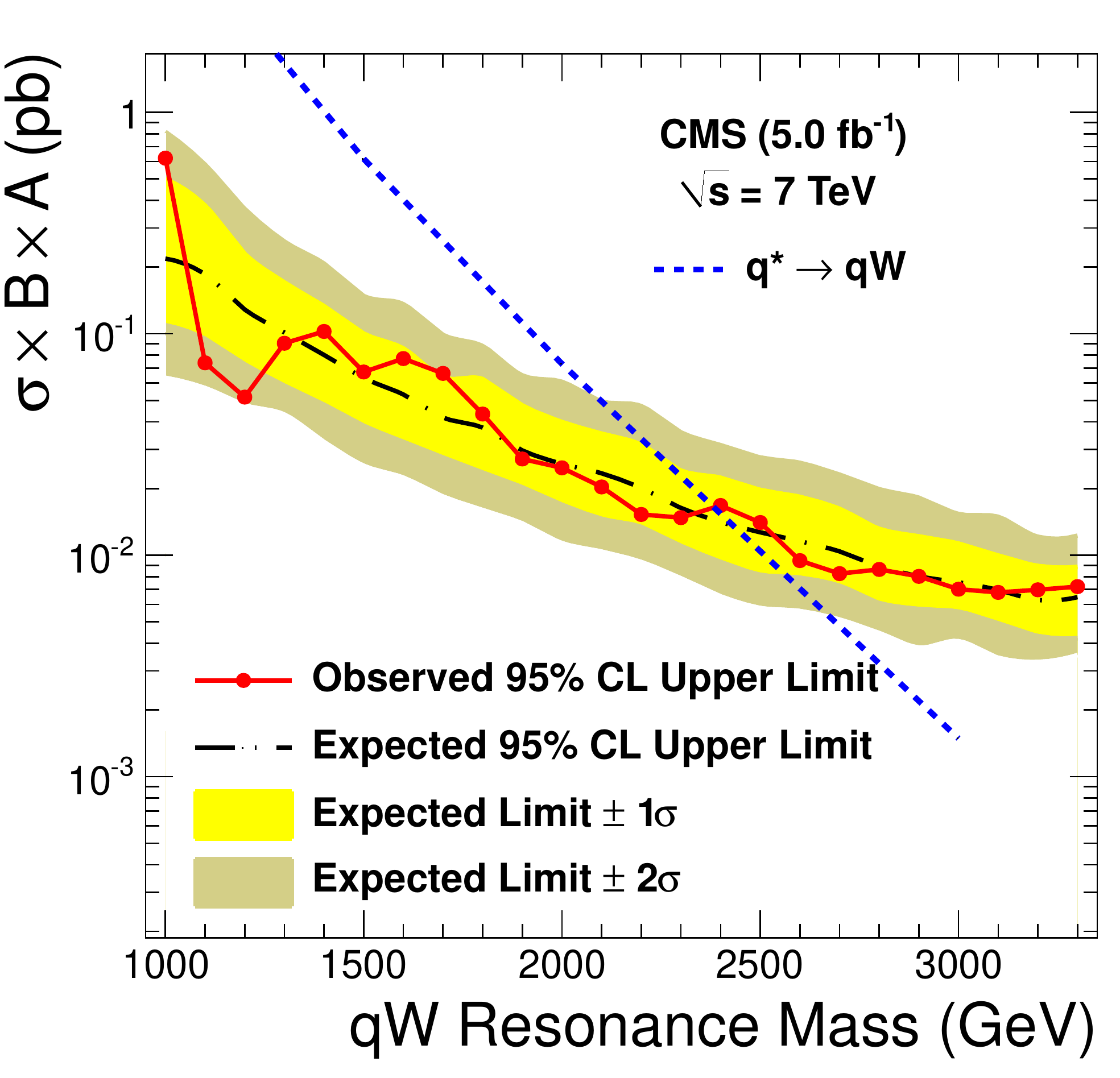}
\includegraphics[width=0.43\textwidth]{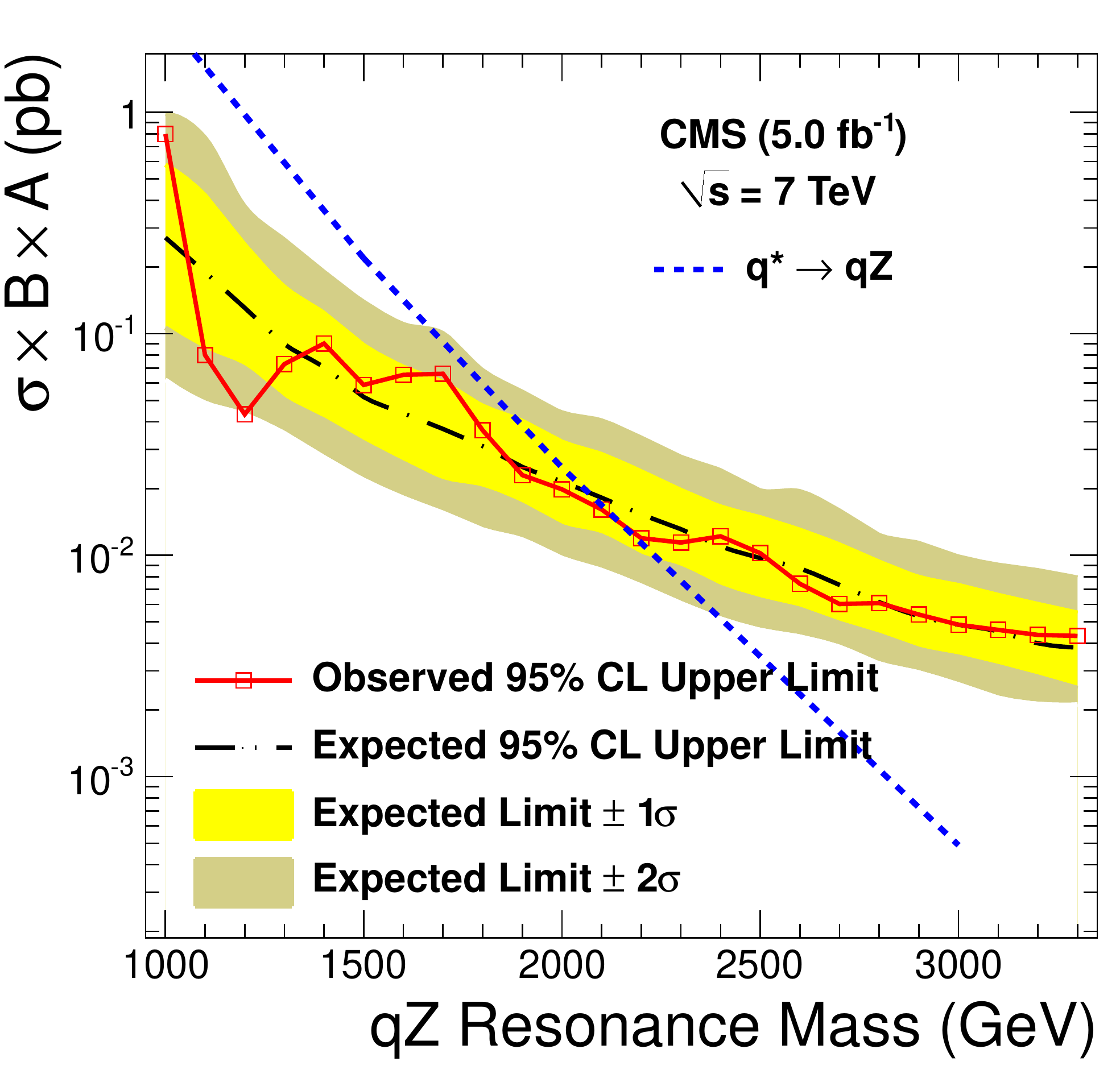}\\
\includegraphics[width=0.43\textwidth]{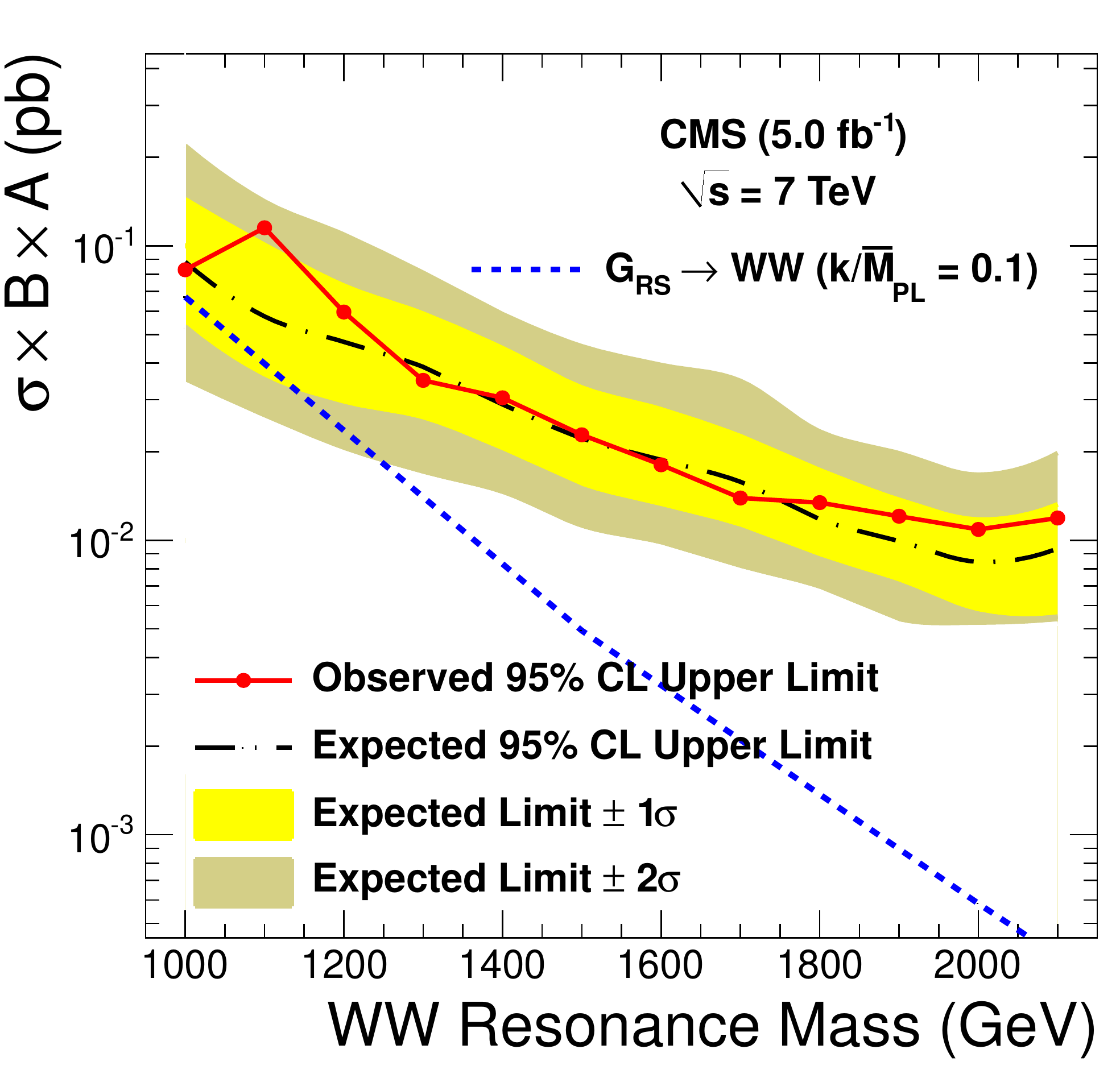}
\includegraphics[width=0.43\textwidth]{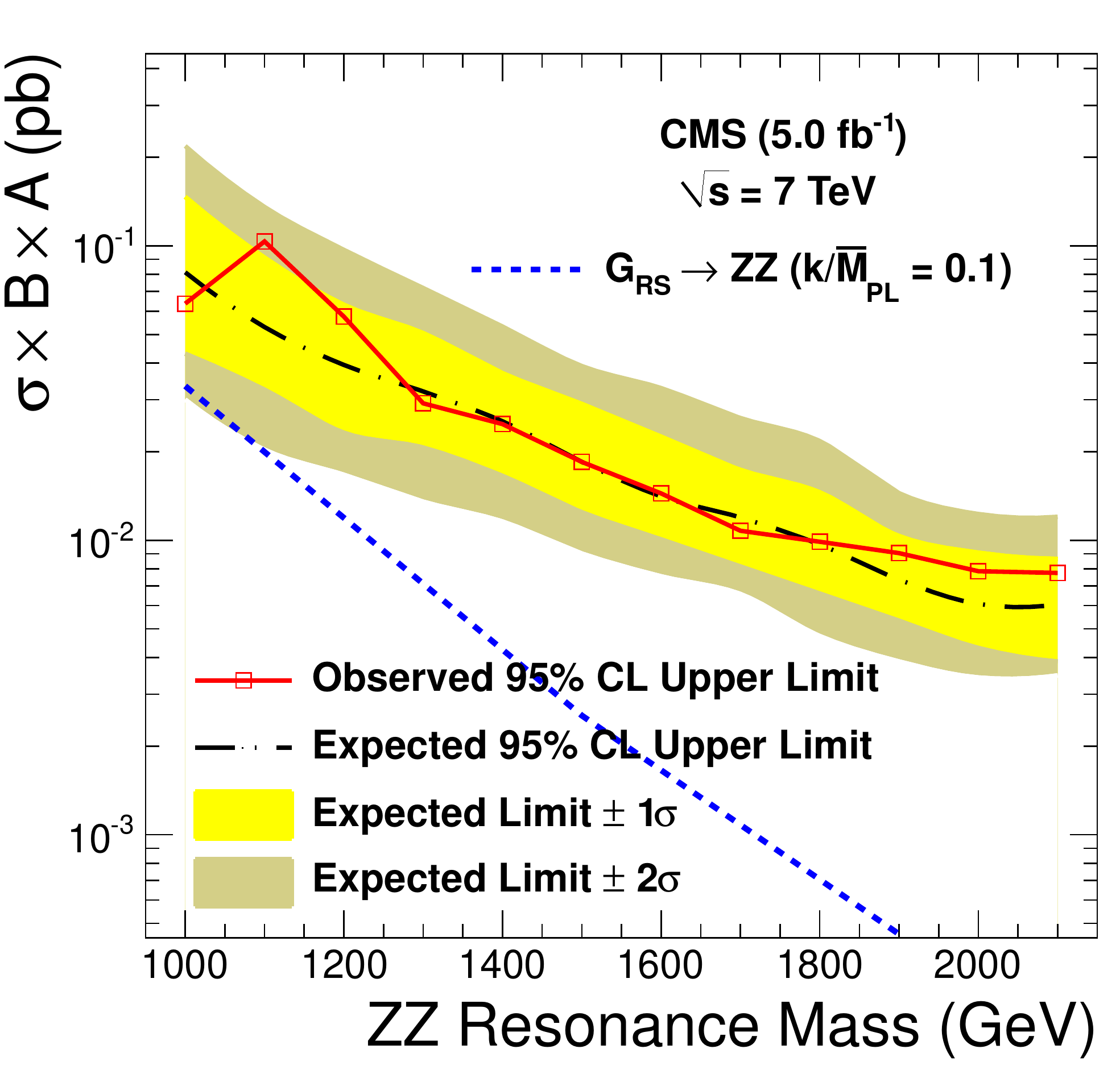}\\
\includegraphics[width=0.43\textwidth]{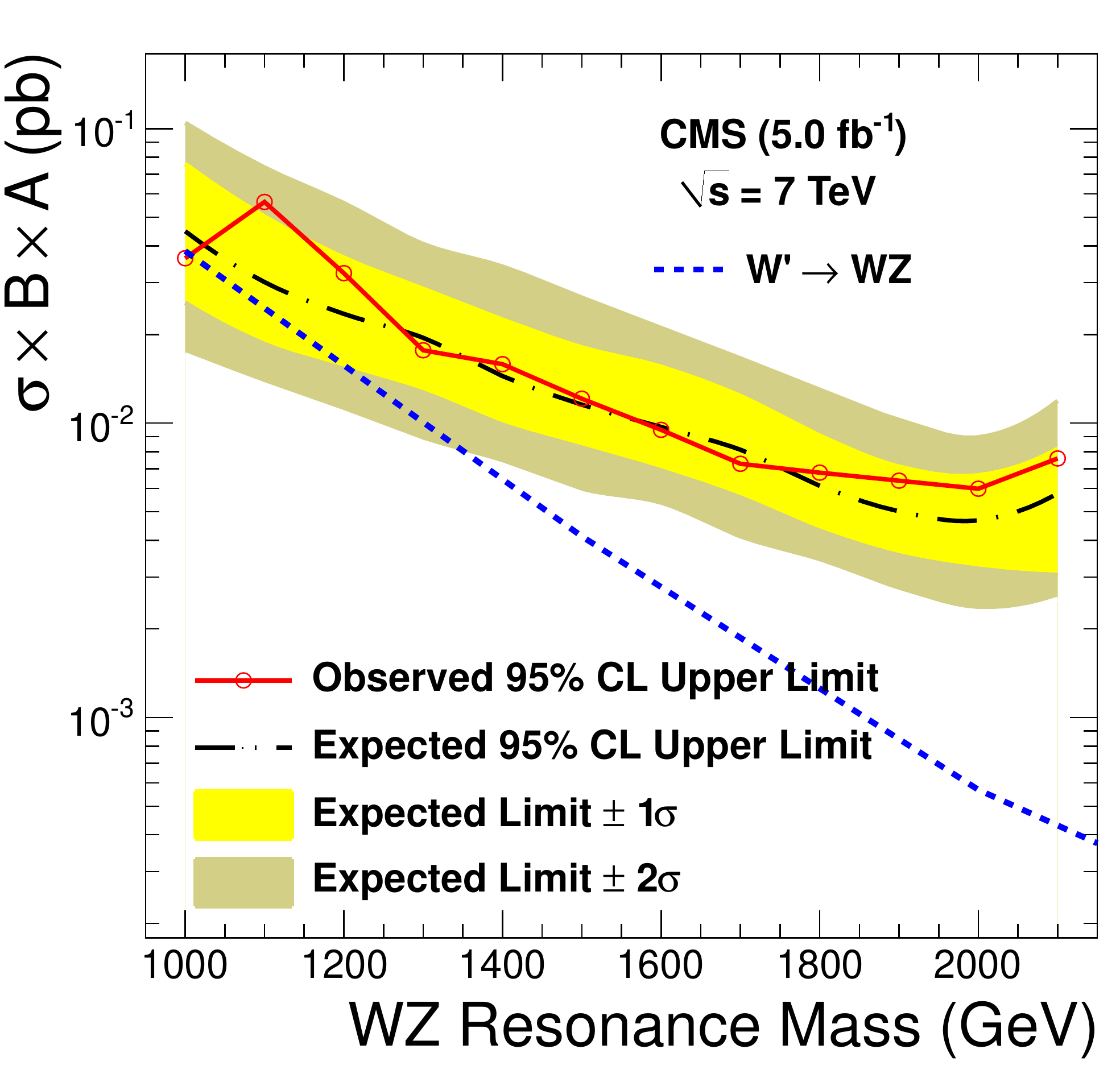}
\end{center}
\caption{Expected and observed limits for qW (top-left), qZ (top-right), WW (center-left), ZZ (center-right) and WZ (bottom) resonances.
  Here, B $\times$ A in the vertical axis label
  contains the branching fraction of $\GRS \to \PW\PW/\cPZ\cPZ \to \text{2 jets}$ or
  $\cPq^* \to \cPq\PW/\cPq\cPZ \to \text{2 jets}$, as well as the acceptance for reconstructing
  the jets in $|\eta| < 2.5$, $|\Delta\eta|<1.3$.
  The predicted cross sections as a function of resonance mass for the considered benchmark models are overlaid.}
\label{fig:Vtagresults}
\end{figure*}

\section{Results}
\label{results}

Figure~\ref{fig:Vtagresults} shows the 95\% CL cross section upper limits
derived from the single and double $\PW/\cPZ$-tagged event samples.
The predicted cross sections as a function of resonance mass for the considered benchmark models are overlaid.
A 95\% CL lower limit is set on the mass of
excited quark resonances decaying into qW (qZ) at \qWlimit (\qZlimit ),
whereas a limit of \qWexplimit (\qZexplimit ) is expected.
These are the most stringent limits in the qW and qZ final states to date.
The sensitivity of our measurement with the present dataset is not
sufficient to extract substantive mass limits on the \GRS with
$k/\MPl=0.1$ nor the heavy SM-like $\PWpr$ boson,
but the cross section limits are the most stringent in the fully hadronic final state to date.
Comparing to the cross section limits on \GRS decays to ZZ in the corresponding semileptonic final states~\cite{CMSZZPAS2,ATLASZZPAPER,CDFZZPAPER},
this analysis sets a stronger cross section limit above a \GRS mass of roughly 1.4\TeVcc .
The cross section limits on final states with ZZ are stronger than
those with WW because the efficiency for tagging a Z is larger than that for a W.
The predicted cross section for WW is twice as large as for ZZ in the \GRS model because of charge-conjugation.

The dominant source of systematic uncertainty is the jet energy
scale. Removing all systematic uncertainties on the single and double
tagged searches would decrease the upper limit on the cross section by
less than 20\% at all values of resonance mass,
which would translate into a change on resonances mass limits of roughly 3\%.
\section{Summary}
\label{sec:conclusions}

A data sample corresponding to an integrated luminosity
of \intlumi collected in pp collisions at $\sqrt{s}=7$\TeVcc with the CMS detector
was used to measure the
$\PW/\cPZ$-tagged dijet mass spectrum using the two leading jets within the
pseudorapidity range $|\eta| < 2.5$ and with pseudorapidity separation
$|\Delta\eta| < 1.3$.
By suppressing the QCD background with the selection of candidates for vector bosons decaying to hadrons within jets,
no evidence was found for new
particle production in the $\PW/\cPZ$-tagged dijet spectrum.
A 95\% CL lower limit is set on the mass of excited quark resonances decaying into
qW (qZ) at \qWlimit (\qZlimit ) and upper limits on the cross section
for resonances decaying to qW, qZ, WW, WZ, or ZZ final states.
These are the most stringent limits in the qW and qZ final states to date.

\section*{Acknowledgements}

We congratulate our colleagues in the CERN accelerator departments for the excellent performance of the LHC and thank the technical and administrative staffs at CERN and at other CMS institutes for their contributions to the success of the CMS effort. In addition, we gratefully acknowledge the computing centres and personnel of the Worldwide LHC Computing Grid for delivering so effectively the computing infrastructure essential to our analyses. Finally, we acknowledge the enduring support for the construction and operation of the LHC and the CMS detector provided by the following funding agencies: BMWF and FWF (Austria); FNRS and FWO (Belgium); CNPq, CAPES, FAPERJ, and FAPESP (Brazil); MEYS (Bulgaria); CERN; CAS, MoST, and NSFC (China); COLCIENCIAS (Colombia); MSES (Croatia); RPF (Cyprus); MoER, SF0690030s09 and ERDF (Estonia); Academy of Finland, MEC, and HIP (Finland); CEA and CNRS/IN2P3 (France); BMBF, DFG, and HGF (Germany); GSRT (Greece); OTKA and NKTH (Hungary); DAE and DST (India); IPM (Iran); SFI (Ireland); INFN (Italy); NRF and WCU (Republic of Korea); LAS (Lithuania); CINVESTAV, CONACYT, SEP, and UASLP-FAI (Mexico); MSI (New Zealand); PAEC (Pakistan); MSHE and NSC (Poland); FCT (Portugal); JINR (Armenia, Belarus, Georgia, Ukraine, Uzbekistan); MON, RosAtom, RAS and RFBR (Russia); MSTD (Serbia); SEIDI and CPAN (Spain); Swiss Funding Agencies (Switzerland); NSC (Taipei); ThEPCenter, IPST and NSTDA (Thailand); TUBITAK and TAEK (Turkey); NASU (Ukraine); STFC (United Kingdom); DOE and NSF (USA).

\bibliography{auto_generated}   

\providecommand{\href}[2]{#2}\begingroup\raggedright\begin{thebibliography}{10}%
\makeatletter
\providecommand{\hrefCMSnoop }[0]{\@secondoftwo}%
\makeatother
\providecommand{\doi}{\texttt{doi:}\begingroup \urlstyle{tt}\Url}

\bibitem{Anchordoqui:2008di}
L.~A. Anchordoqui\hrefCMSnoop {} { {et~al.}, ``Jet signals for low mass strings
  at the {L}arge {H}adron {C}ollider'',} \textit{ Phys. Rev. Lett.} \textbf{
  100} (2008) 171603,
  \href{http://dx.doi.org/10.1103/Physrevlett.100.171603}{\doi{10.1103/Physrevlett.100.171603}},
  \href{http://www.arXiv.org/abs/0712.0386}{\texttt{ arXiv:0712.0386}}.

\bibitem{Cullen:2000ef}
\hrefCMSnoop {} {S.~Cullen, M.~Perelstein, and M.~E. Peskin, ``{TeV} strings
  and collider probes of large extra dimensions'',} \textit{ Phys. Rev. D}
  \textbf{ 62} (2000) 055012,
  \href{http://dx.doi.org/10.1103/PhysRevD.62.055012}{\doi{10.1103/PhysRevD.62.055012}},
  \href{http://www.arXiv.org/abs/hep-ph/0001166}{\texttt{
  arXiv:hep-ph/0001166}}.

\bibitem{ref_diquark}
\hrefCMSnoop {} {J.~L. Hewett and T.~G. Rizzo, ``Low-energy phenomenology of
  superstring-inspired {$E_6$} models'',} \textit{ Phys. Rept.} \textbf{ 183}
  (1989) 193,
  \href{http://dx.doi.org/10.1016/0370-1573(89)90071-9}{\doi{10.1016/0370-1573(89)90071-9}}.

\bibitem{ref_qstar}
\hrefCMSnoop {} {U.~Baur, I.~Hinchliffe, and D.~Zeppenfeld, ``Excited Quark
  Production at Hadron Colliders'',} \textit{ Int. J. Mod. Phys. A} \textbf{ 2}
  (1987) 1285,
  \href{http://dx.doi.org/10.1142/S0217751X87000661}{\doi{10.1142/S0217751X87000661}}.

\bibitem{Baur:1989kv}
\hrefCMSnoop {} {U.~Baur, M.~Spira, and P.~M. Zerwas, ``Excited Quark and
  Lepton Production at Hadron Colliders'',} \textit{ Phys. Rev. D} \textbf{ 42}
  (1990) 815,
\href{http://dx.doi.org/10.1103/PhysRevD.42.815}{\doi{10.1103/PhysRevD.42.815}}.

\bibitem{ref_axi}
\hrefCMSnoop {} {P.~H. Frampton and S.~L. Glashow, ``Chiral color: An
  alternative to the standard model'',} \textit{ Phys. Lett. B} \textbf{ 190}
  (1987) 157,
  \href{http://dx.doi.org/10.1016/0370-2693(87)90859-8}{\doi{10.1016/0370-2693(87)90859-8}}.

\bibitem{ref_coloron}
\hrefCMSnoop {} {E.~H. Simmons, ``Coloron phenomenology'',} \textit{ Phys. Rev.
  D} \textbf{ 55} (1997) 1678,
  \href{http://dx.doi.org/10.1103/PhysRevD.55.1678}{\doi{10.1103/PhysRevD.55.1678}},
  \href{http://www.arXiv.org/abs/hep-ph/9608269}{\texttt{
  arXiv:hep-ph/9608269}}.

\bibitem{ref_gauge}
E.~Eichten\hrefCMSnoop {} { {et~al.}, ``Supercollider physics'',} \textit{ Rev.
  Mod. Phys.} \textbf{ 56} (1984) 579,
  \href{http://dx.doi.org/10.1103/RevModPhys.56.579}{\doi{10.1103/RevModPhys.56.579}}.

\bibitem{ref_rsg}
\hrefCMSnoop {} {L.~Randall and R.~Sundrum, ``An alternative to
  compactification'',} \textit{ Phys. Rev. Lett.} \textbf{ 83} (1999) 4690,
  \href{http://dx.doi.org/10.1103/PhysRevLett.83.4690}{\doi{10.1103/PhysRevLett.83.4690}}.

\bibitem{exo12094}
\hrefCMSnoop {} {{CMS Collaboration}, ``Search for narrow resonances and
  quantum black holes in inclusive and b-tagged dijet mass spectra from pp
  collisions at $\sqrt{s}$ = 7 {TeV}'',} (2012).
  \href{http://www.arXiv.org/abs/1210.2387}{\texttt{ arXiv:1210.2387}}.
Submitted to JHEP.

\bibitem{ATLASexcitedPAS}
\hrefCMSnoop {} {{ATLAS Collaboration}, ``{ATLAS} search for new phenomena in
  dijet mass and angular distributions using pp collisions at $\sqrt{s}$ = 7
  {TeV}'',} (2012). \href{http://www.arXiv.org/abs/1210.1718}{\texttt{
  arXiv:1210.1718}}.
Submitted to JHEP.

\bibitem{catop_cms}
\href {http://cdsweb.cern.ch/record/1194489} {{CMS Collaboration}, ``{A
  Cambridge-Aachen (C-A) based Jet Algorithm for boosted top-jet tagging}'',}
  CMS Physics Analysis Summary CMS-PAS-JME-09-001, (2009).

\bibitem{rs1}
\hrefCMSnoop {} {L.~Randall and R.~Sundrum, ``A large mass hierarchy from a
  small extra dimension'',} \textit{ Phys. Rev. Lett.} \textbf{ 83} (1999)
  3370,
  \href{http://dx.doi.org/10.1103/PhysRevLett.83.3370}{\doi{10.1103/PhysRevLett.83.3370}},
  \href{http://www.arXiv.org/abs/hep-ph/9905221}{\texttt{
  arXiv:hep-ph/9905221}}.

\bibitem{CDFexcitedPAPER}
\hrefCMSnoop {} {{ CDF} Collaboration, ``Search for excited quarks in
  $p\bar{p}$ collisions at $\sqrt{s}$ = 1.8 {TeV}'',} \textit{ Phys. Rev.
  Lett.} \textbf{ 72} (1994) 3004,
  \href{http://dx.doi.org/10.1103/PhysRevLett.72.3004}{\doi{10.1103/PhysRevLett.72.3004}}.

\bibitem{D0excitedPAPER}
\hrefCMSnoop {} {{ D0} Collaboration, ``Search for a heavy resonance decaying
  into a {Z+ jet} final state in $p\bar{p}$ collisions at $\sqrt{s}$ = 1.96
  {TeV} using the {D0} detector'',} \textit{ Phys. Rev. D} \textbf{ 74} (2006)
  011104,
  \href{http://dx.doi.org/10.1103/PhysRevD.74.011104}{\doi{10.1103/PhysRevD.74.011104}},
  \href{http://www.arXiv.org/abs/hep-ex/0606018}{\texttt{
  arXiv:hep-ex/0606018}}.

\bibitem{CMSqZPAS}
\hrefCMSnoop {} {{CMS Collaboration}, ``Search for anomalous production of
  highly boosted {Z} bosons decaying to dimuons in pp collisions at $\sqrt{s}$
  = 7 {TeV}'',} (2012). \href{http://www.arXiv.org/abs/1210.0867}{\texttt{
  arXiv:1210.0867}}.
Submitted to Phys. Lett. B.

\bibitem{GravitonWWZZ1}
K.~Agashe\hrefCMSnoop {} { {et~al.}, ``Warped Gravitons at the {LHC} and
  Beyond'',} \textit{ Phys. Rev. D} \textbf{ 76} (2007) 036006,
  \href{http://dx.doi.org/10.1103/PhysRevD.76.036006}{\doi{10.1103/PhysRevD.76.036006}},
  \href{http://www.arXiv.org/abs/hep-ph/0701186}{\texttt{
  arXiv:hep-ph/0701186}}.

\bibitem{GravitonWWZZ2}
\hrefCMSnoop {} {O.~Antipin, D.~Atwood, and A.~Soni, ``Search for {RS}
  gravitons via $W_L W_L$ decays'',} \textit{ Phys. Lett. B} \textbf{ 666}
  (2008) 155,
  \href{http://dx.doi.org/10.1016/j.physletb.2008.07.009}{\doi{10.1016/j.physletb.2008.07.009}},
  \href{http://www.arXiv.org/abs/0711.3175}{\texttt{ arXiv:0711.3175}}.

\bibitem{GravitonWWZZ3}
\hrefCMSnoop {} {O.~Antipin and A.~Soni, ``Towards establishing the spin of
  warped gravitons'',} \textit{ JHEP} \textbf{ 10} (2008) 018,
  \href{http://dx.doi.org/10.1088/1126-6708/2008/10/018}{\doi{10.1088/1126-6708/2008/10/018}},
  \href{http://www.arXiv.org/abs/0806.3427}{\texttt{ arXiv:0806.3427}}.

\bibitem{CMSZZPAS2}
\hrefCMSnoop {} {{CMS Collaboration}, ``Search for a narrow spin-2 resonance
  decaying to a pair of {Z} vector bosons in the semileptonic final state'',}
  (2012). \href{http://www.arXiv.org/abs/1209.3807}{\texttt{ arXiv:1209.3807}}.
Submitted to Phys. Lett. B.

\bibitem{ATLASZZPAPER}
\hrefCMSnoop {} {{ ATLAS} Collaboration, ``Search for new particles decaying to
  {ZZ} using final states with leptons and jets with the {ATLAS} detector in
  $\sqrt{s}$ = 7 {TeV} proton-proton collisions'',} \textit{ Phys. Lett. B}
  \textbf{ 712} (2012) 331,
  \href{http://dx.doi.org/10.1016/j.physletb.2012.05.020}{\doi{10.1016/j.physletb.2012.05.020}},
  \href{http://www.arXiv.org/abs/1203.0718}{\texttt{ arXiv:1203.0718}}.

\bibitem{CDFZZPAPER}
\hrefCMSnoop {} {{ CDF} Collaboration, ``Search for high-mass resonances
  decaying into {ZZ} in $p\bar{p}$ collisions at $\sqrt{s}$ = 1.96 {TeV}'',}
  \textit{ Phys. Rev. D} \textbf{ 85} (2012) 012008,
  \href{http://dx.doi.org/10.1103/PhysRevD.85.012008}{\doi{10.1103/PhysRevD.85.012008}},
  \href{http://www.arXiv.org/abs/1111.3432}{\texttt{ arXiv:1111.3432}}.

\bibitem{CMSwprimePAS}
\hrefCMSnoop {} {{ CMS} Collaboration, ``Search for leptonic decays of {\PWpr}
  bosons in pp collisions at {$\sqrt{s}=7\TeV$}'',} \textit{ JHEP} \textbf{ 08}
  (2012) 023,
  \href{http://dx.doi.org/10.1007/JHEP08(2012)023}{\doi{10.1007/JHEP08(2012)023}}.

\bibitem{ATLASwprimePAPER}
\hrefCMSnoop {} {{ ATLAS} Collaboration, ``Search for a heavy gauge boson
  decaying to a charged lepton and a neutrino in 1 fb$^{-1}$ of pp collisions
  at $\sqrt{s}$ = 7 {TeV} using the {ATLAS} detector'',} \textit{ Phys. Lett.
  B} \textbf{ 705} (2011) 28,
  \href{http://dx.doi.org/10.1016/j.physletb.2011.09.093}{\doi{10.1016/j.physletb.2011.09.093}},
  \href{http://www.arXiv.org/abs/1108.1316}{\texttt{ arXiv:1108.1316}}.

\bibitem{CMSwprimeWZPAS}
\hrefCMSnoop {} {{ CMS} Collaboration, ``Search for a {\PWpr\ or
  $\rho_\mathrm{TC}$ decaying to $\PW\cPZ$ in $\Pp\Pp$ collisions at
  $\sqrt{s}=7$\TeV}'',} \textit{ Phys. Rev. Lett.} \textbf{ 109} (2012) 141801,
  \href{http://dx.doi.org/10.1103/PhysRevLett.109.141801}{\doi{10.1103/PhysRevLett.109.141801}}.

\bibitem{ATLASwprimeWZPAS}
\hrefCMSnoop {} {{ ATLAS} Collaboration, ``Search for resonant {WZ} production
  in the {WZ} to l nu l' l' channel in $\sqrt{s}$ = 7 {TeV} pp collisions with
  the {ATLAS} detector'',} \textit{ Phys. Rev. D} \textbf{ 85} (2012) 112012,
  \href{http://dx.doi.org/10.1103/PhysRevD.85.112012}{\doi{10.1103/PhysRevD.85.112012}},
  \href{http://www.arXiv.org/abs/1204.1648}{\texttt{ arXiv:1204.1648}}.

\bibitem{:2008zzk}
\hrefCMSnoop {} {{ CMS} Collaboration, ``The {CMS} experiment at the {CERN}
  {LHC}'',} \textit{ JINST} \textbf{ 03} (2008) S08004,
\href{http://dx.doi.org/10.1088/1748-0221/3/08/S08004}{\doi{10.1088/1748-0221/3/08/S08004}}.

\bibitem{particleflow}
\href {http://cdsweb.cern.ch/record/1194487} {{CMS Collaboration},
  ``Particle--Flow Event Reconstruction in {CMS} and Performance for Jets,
  Taus, and {\MET}'',} CMS Physics Analysis Summary CMS-PAS-PFT-09-001, (2009).

\bibitem{pythia}
\hrefCMSnoop {} {T.~Sj{\"o}strand, S.~Mrenna, and P.~Skands, ``{PYTHIA} 6.4
  physics and manual'',} \textit{ JHEP} \textbf{ 05} (2006) 026,
  \href{http://dx.doi.org/10.1088/1126-6708/2006/05/026}{\doi{10.1088/1126-6708/2006/05/026}},
\href{http://www.arXiv.org/abs/hep-ph/0603175}{\texttt{ arXiv:hep-ph/0603175}}.

\bibitem{herwig}
M.~B{\"a}hr\hrefCMSnoop {} { {et~al.}, ``{Herwig++} physics and manual'',}
  \textit{ Eur. Phys. J. C} \textbf{ 58} (2008) 639,
  \href{http://dx.doi.org/10.1140/epjc/s10052-008-0798-9}{\doi{10.1140/epjc/s10052-008-0798-9}},
\href{http://www.arXiv.org/abs/0803.0883}{\texttt{ arXiv:0803.0883}}.

\bibitem{cteq}
J.~Pumplin\hrefCMSnoop {} { {et~al.}, ``New generation of parton distributions
  with uncertainties from global {QCD} analysis'',} \textit{ JHEP} \textbf{ 07}
  (2002) 012,
  \href{http://dx.doi.org/10.1088/1126-6708/2002/07/012}{\doi{10.1088/1126-6708/2002/07/012}},
  \href{http://www.arXiv.org/abs/hep-ph/0201195}{\texttt{
  arXiv:hep-ph/0201195}}.

\bibitem{mrst}
A.~D. Martin\hrefCMSnoop {} { {et~al.}, ``{MRST2001}: partons and alpha\_S from
  precise deep inelastic scattering and {Tevatron} jet data'',} \textit{ Eur.
  Phys. J. C} \textbf{ 23} (2002) 73,
  \href{http://dx.doi.org/10.1007/s100520100842}{\doi{10.1007/s100520100842}},
  \href{http://www.arXiv.org/abs/hep-ph/0110215}{\texttt{
  arXiv:hep-ph/0110215}}.

\bibitem{bib_tunez1}
\hrefCMSnoop {} {R.~Field, ``Early {LHC} Underlying Event Data -- Findings and
  Surprises'',} in \textit{ Proceeedings of the Hadron Collider Physics
  Symposium 2010}.
\newblock 2010.
\newblock
\href{http://www.arXiv.org/abs/1010.3558}{\texttt{ arXiv:1010.3558}}.
\newblock

\bibitem{resonanceshape}
\hrefCMSnoop {} {{CMS Collaboration}, ``Search for {Randall-Sundrum} Graviton
  Excitations in the {CMS} Experiment'',} (2002).
  \href{http://www.arXiv.org/abs/hep-ex/0207061}{\texttt{
  arXiv:hep-ex/0207061}}.

\bibitem{refGEANT}
\hrefCMSnoop {} {{ GEANT4} Collaboration, ``{GEANT4}---a simulation toolkit'',}
  \textit{ Nucl. Instrum. Meth. A} \textbf{ 506} (2003) 250,
  \href{http://dx.doi.org/10.1016/S0168-9002(03)01368-8}{\doi{10.1016/S0168-9002(03)01368-8}}.

\bibitem{ktalg}
\hrefCMSnoop {} {M.~Cacciari, G.~P. Salam, and G.~Soyez, ``The anti-$k_t$ jet
  clustering algorithm'',} \textit{ JHEP} \textbf{ 04} (2008) 063,
  \href{http://dx.doi.org/10.1088/1126-6708/2008/04/063}{\doi{10.1088/1126-6708/2008/04/063}},
\href{http://www.arXiv.org/abs/0802.1189}{\texttt{ arXiv:0802.1189}}.

\bibitem{CAaachen}
\hrefCMSnoop {} {M.~Wobisch and T.~Wengler, ``Hadronization corrections to jet
  cross sections in deep-inelastic scattering'',} (1998).
\href{http://www.arXiv.org/abs/hep-ph/9907280}{\texttt{ arXiv:hep-ph/9907280}}.

\bibitem{CAcambridge}
Y.~L. Dokshitzer\hrefCMSnoop {} { {et~al.}, ``Better jet clustering
  algorithms'',} \textit{ JHEP} \textbf{ 08} (1997) 001,
  \href{http://dx.doi.org/10.1088/1126-6708/1997/08/001}{\doi{10.1088/1126-6708/1997/08/001}},
\href{http://www.arXiv.org/abs/hep-ph/9707323}{\texttt{ arXiv:hep-ph/9707323}}.

\bibitem{fastjet1}
\hrefCMSnoop {} {M.~Cacciari and G.~P. Salam, ``Dispelling the $N^3$ myth for
  the $k_t$ jet-finder'',} \textit{ Phys. Lett. B} \textbf{ 641} (2006) 57,
  \href{http://dx.doi.org/10.1016/j.physletb.2006.08.037}{\doi{10.1016/j.physletb.2006.08.037}},
\href{http://www.arXiv.org/abs/hep-ph/0512210}{\texttt{ arXiv:hep-ph/0512210}}.

\bibitem{fastjet}
\hrefCMSnoop {} {M.~Cacciari, G.~P. Salam, and G.~Soyez, ``FastJet User
  Manual'',} \textit{ Eur. Phys. J. C} \textbf{ 72} (2012) 1896,
  \href{http://dx.doi.org/10.1140/epjc/s10052-012-1896-2}{\doi{10.1140/epjc/s10052-012-1896-2}},
  \href{http://www.arXiv.org/abs/1111.6097}{\texttt{ arXiv:1111.6097}}.

\bibitem{jetarea_fastjet}
\hrefCMSnoop {} {M.~Cacciari, G.~P. Salam, and G.~Soyez, ``The Catchment Area
  of Jets'',} \textit{ JHEP} \textbf{ 04} (2008) 005,
  \href{http://dx.doi.org/10.1088/1126-6708/2008/04/005}{\doi{10.1088/1126-6708/2008/04/005}},
\href{http://www.arXiv.org/abs/0802.1188}{\texttt{ arXiv:0802.1188}}.

\bibitem{jetarea_fastjet_pu}
\hrefCMSnoop {} {M.~Cacciari and G.~P. Salam, ``Pileup subtraction using jet
  areas'',} \textit{ Phys. Lett. B} \textbf{ 659} (2008) 119,
  \href{http://dx.doi.org/10.1016/j.physletb.2007.09.077}{\doi{10.1016/j.physletb.2007.09.077}},
\href{http://www.arXiv.org/abs/0707.1378}{\texttt{ arXiv:0707.1378}}.

\bibitem{JME-JINST}
\hrefCMSnoop {} {{ CMS} Collaboration, ``Determination of jet energy
  calibration and transverse momentum resolution in {CMS}'',} \textit{ JINST}
  \textbf{ 6} (2011) P11002,
  \href{http://dx.doi.org/10.1088/1748-0221/6/11/P11002}{\doi{10.1088/1748-0221/6/11/P11002}}.

\bibitem{CMSexcitedPAPER}
\hrefCMSnoop {} {{ CMS} Collaboration, ``Search for resonances in the dijet
  mass spectrum from 7~{TeV} pp collisions at {CMS}'',} \textit{ Phys. Lett. B}
  \textbf{ 704} (2011) 123,
  \href{http://dx.doi.org/10.1016/j.physletb.2011.09.015}{\doi{10.1016/j.physletb.2011.09.015}}.

\bibitem{topwtag_pas}
\href {http://cdsweb.cern.ch/record/1333700} {{CMS Collaboration}, ``Study of
  Jet Substructure in pp Collisions at 7 {TeV} in {CMS}'',} CMS Physics
  Analysis Summary CMS-PAS-JME-10-013, (2010).

\bibitem{jetpruning1}
\hrefCMSnoop {} {S.~D. Ellis, C.~K. Vermilion, and J.~R. Walsh, ``Techniques
  for improved heavy particle searches with jet substructure'',} \textit{ Phys.
  Rev. D} \textbf{ 80} (2009) 051501,
  \href{http://dx.doi.org/10.1103/PhysRevD.80.051501}{\doi{10.1103/PhysRevD.80.051501}},
  \href{http://www.arXiv.org/abs/0903.5081}{\texttt{ arXiv:0903.5081}}.

\bibitem{jetpruning2}
\hrefCMSnoop {} {S.~D. Ellis, C.~K. Vermilion, and J.~R. Walsh, ``Recombination
  Algorithms and Jet Substructure: Pruning as a Tool for Heavy Particle
  Searches'',} \textit{ Phys. Rev. D} \textbf{ 81} (2010) 094023,
  \href{http://dx.doi.org/10.1103/PhysRevD.81.094023}{\doi{10.1103/PhysRevD.81.094023}},
  \href{http://www.arXiv.org/abs/0912.0033}{\texttt{ arXiv:0912.0033}}.

\bibitem{boostedhiggs}
J.~M. Butterworth\hrefCMSnoop {} { {et~al.}, ``Jet substructure as a new Higgs
  search channel at the {LHC}'',} \textit{ Phys. Rev. Lett.} \textbf{ 100}
  (2008) 242001,
  \href{http://dx.doi.org/10.1103/PhysRevLett.100.242001}{\doi{10.1103/PhysRevLett.100.242001}},
\href{http://www.arXiv.org/abs/0802.2470}{\texttt{ arXiv:0802.2470}}.

\bibitem{madgraph}
J.~Alwall\hrefCMSnoop {} { {et~al.}, ``{MadGraph/MadEvent v4}: the new web
  generation'',} \textit{ JHEP} \textbf{ 09} (2007) 028,
  \href{http://dx.doi.org/10.1088/1126-6708/2007/09/028}{\doi{10.1088/1126-6708/2007/09/028}},
\href{http://www.arXiv.org/abs/0706.2334}{\texttt{ arXiv:0706.2334}}.

\bibitem{CMSttbar}
\hrefCMSnoop {} {{ CMS} Collaboration, ``Search for anomalous $t\bar{t}$
  production in the highly-boosted all-hadronic final state'',} \textit{ JHEP}
  \textbf{ 09} (2012) 029,
  \href{http://dx.doi.org/10.1007/JHEP09(2012)029}{\doi{10.1007/JHEP09(2012)029}},
  \href{http://www.arXiv.org/abs/1204.2488}{\texttt{ arXiv:1204.2488}}.

\bibitem{cmsdijet}
\hrefCMSnoop {} {{ CMS} Collaboration, ``Search for resonances in the dijet
  mass spectrum from 7 {TeV} pp collisions at {CMS}'',} \textit{ Phys. Lett. B}
  \textbf{ 704} (2011) 123,
  \href{http://dx.doi.org/10.1016/j.physletb.2011.09.015}{\doi{10.1016/j.physletb.2011.09.015}},
  \href{http://www.arXiv.org/abs/1107.4771}{\texttt{ arXiv:1107.4771}}.

\bibitem{Ftest}
R.~G. Lomax and D.~L. Hahs-Vaughn, ``Statistical Concepts: A Second Course''.
\newblock p.10, Routledge Academic, 2007.

\bibitem{SMP-12-008-PAS}
\href {http://cdsweb.cern.ch/record/1434360} {{CMS Collaboration}, ``Absolute
  Calibration of the Luminosity Measurement at {CMS}: {W}inter 2012 Update'',}
  CMS Physics Analysis Summary CMS-PAS-SMP-12-008, (2012).

\end{thebibliography}\endgroup

\cleardoublepage \appendix\section{The CMS Collaboration \label{app:collab}}\begin{sloppypar}\hyphenpenalty=5000\widowpenalty=500\clubpenalty=5000\textbf{Yerevan Physics Institute,  Yerevan,  Armenia}\\*[0pt]
S.~Chatrchyan, V.~Khachatryan, A.M.~Sirunyan, A.~Tumasyan
\vskip\cmsinstskip
\textbf{Institut f\"{u}r Hochenergiephysik der OeAW,  Wien,  Austria}\\*[0pt]
W.~Adam, E.~Aguilo, T.~Bergauer, M.~Dragicevic, J.~Er\"{o}, C.~Fabjan\cmsAuthorMark{1}, M.~Friedl, R.~Fr\"{u}hwirth\cmsAuthorMark{1}, V.M.~Ghete, J.~Hammer, N.~H\"{o}rmann, J.~Hrubec, M.~Jeitler\cmsAuthorMark{1}, W.~Kiesenhofer, V.~Kn\"{u}nz, M.~Krammer\cmsAuthorMark{1}, I.~Kr\"{a}tschmer, D.~Liko, I.~Mikulec, M.~Pernicka$^{\textrm{\dag}}$, B.~Rahbaran, C.~Rohringer, H.~Rohringer, R.~Sch\"{o}fbeck, J.~Strauss, A.~Taurok, W.~Waltenberger, G.~Walzel, E.~Widl, C.-E.~Wulz\cmsAuthorMark{1}
% ...
\vskip\cmsinstskip
\dag:~Deceased\\
1:~~Also at Vienna University of Technology, Vienna, Austria\\
% ...

\end{sloppypar}
\end{document}